\documentclass[prerint,12pt]{elsarticle}

\usepackage{booktabs}
\usepackage{graphicx}
\usepackage[table,xcdraw]{xcolor}
\usepackage{amssymb}

\newcommand{\tabitem}{~~\llap{\textbullet}~~}
\journal{Computer \& Security}

\begin{document}

\begin{frontmatter}

\title{
Datasets are not Enough: Challenges in Labeling Network Traffic}

\author{Jorge Luis Guerra\fnref{label1}}
\fntext[label1]{jorge.guerra@ingenieria.uncuyo.edu.ar}


\author{Carlos Catania, Eduardo Veas}


\begin{abstract}



In contrast to previous surveys, the present work is not focused on reviewing the datasets used in the network security field. The fact is that many of the available public labeled datasets represent the network behavior just for a particular time period. Given the rate of change in malicious behavior and the serious challenge to label, and maintain these datasets, they become quickly obsolete. Therefore, this work is focused on the analysis of current labeling methodologies applied to network-based data. In the field of network security, the process of labeling a representative network traffic dataset is particularly challenging and costly since very specialized knowledge is required to classify network traces. Consequently, most of the current traffic labeling methods are based on the automatic generation of synthetic network traces, which hides many of the essential aspects necessary for a correct differentiation between normal and malicious behavior. Alternatively, a few other methods incorporate non-experts users in the labeling process of real traffic with the help of visual and statistical tools. However, after conducting an in-depth analysis, it seems that all current methods for labeling suffer from fundamental drawbacks regarding the quality, volume, and speed of the resulting dataset. This lack of consistent methods for continuously generating a representative dataset with an accurate and validated methodology must be addressed by the network security research community. Moreover, a consistent label methodology is a fundamental condition for helping in the acceptance of novel detection approaches based on statistical and machine learning techniques. 


\end{abstract}

\begin{keyword}
Network Security \sep Automatic Labeling \sep Assisted Labeling \sep Datasets \sep Network Traffic 


\end{keyword}

\end{frontmatter}


\section{Introduction and Motivation}
\label{sec:introduction}
A Network Intrusion Detection System (NIDS) is an active process that monitors network traffic to identify security breaches and initiate measures to counteract the type of attack (e.g., spam, information stealing, botnet attacks, among others.). Today's network environments suffer from constant modification and improvements. Therefore, a rapid adaptation by NIDS is necessary if they do not want to become obsolete \cite{Resende2018, Glass-Vanderlan2018}. Consequently, NIDS  based on statistical methods, machine learning, and data mining methods have increased their application in recent years mostly because of their generalization capabilities \cite{Buczak2016, Catania2012}.

However, much of the success of the so-called statistically based NIDS (SNIDS) will depend mostly on the initial model generation and the benchmarking before going into the production network infrastructure \cite{Vasilomanolakis2015}. Both procedures will heavily relies on the quality of labeled datasets used.


Although dataset quality is not precisely defined, several authors \cite{Sharafaldin2018, MaciaFernandez2018} agree that representative and accurate labels are the main two aspects for measuring the quality of a network traffic labeled dataset. A representative labeled dataset should provide all the associated behavioral patterns for malicious and normal network traces. Representativeness is particularly important when labeling network traces from normal users, where timing patterns, frequency of use and work cycle must be precisely included in the dataset. In the case of malicious network traces, the sequence of misuse actions performed on the network and their periodicity patterns are examples of representative information. On the other hand, an accurate label should be assigned only to those portions of a network trace containing the behavior of interest. A mislabeled and underrepresented dataset will have direct consequences on the performance of any model generated from the data.

Several aspects can be studied during the generation of labeled datasets for the network security field, such as the mechanism used during the traffic capture \cite{Hofstede, Kumar, Sun, ZARPELAO201725, Pham, Costa}, the subsequent cleaning process \cite{Tesfahun2013, Yueai}, the method of feature extraction \cite{Wheelus, Haddadi, McKeown, Cugola2012}, and the strategy for labeling the network traces, among others. In the particular case of the labeling strategy, it is possible to analyze the process as a simple detection/classification problem in which a given network traffic event is classified as normal or malicious. However, there are meaningful differences in the process of traffic labeling compared with a conventional traffic detection process. These can be framed under the following aspects:

\begin{itemize}

\item \textbf{Timing:} in the labeling process, there is no need to perform the detection in a particular time frame. During the labeling process, the security analyst (automated system or expert user) can take the required time to confirm the potential anomaly or misuse. \textbf{(D1)}

\item \textbf{Relevance:} a false positive is not as crucial for a real-time detection system as it is for a labeled data set creation system. A false positive is an inconvenience to the user during real-time analysis. For the labeling process, however, it merely represents part of the noise that might occur in the resulting dataset. \textbf{(D2)}

\item\textbf{Qualitative:} the focus of the labeling process is to get a set of accurate labels representing the most significant characteristics of the network. The more representative is data, 
the better will be the resulting model for performing detection.
As an example, a labeled dataset with a considerable set of confirmed malicious network traces coming from a unique source and following the same pattern could be easy to predict. However, it might not be useful for generating a proper detection model. (\textbf{D3)}

\item \textbf{Scope:} the scope between the detection problem and the labeling of network traces are often different. Usually, network security datasets are created with a particular scope in mind. On the other hand, when performing real-time detection on real network environments, the detection of malicious traffic does not restrict between network traces. It has the task of classifying all traffic. \textbf{(D4)}

\item \textbf{Economic:} the labeling process has no immediate economic consequences. In other words, when confronted with an undetected malicious network trace, in general, there is no consequence beyond inadequate data for the construction of a statistical prediction model. While in the case of a operational detection system, the non-recognition of malicious behavior can cause important losses to the organization. \textbf{(D5)}

\end{itemize}

Over the past 20 years, several methods have been developed to address the problem of labeling applied to network data sets. One of the most widely used methods has been using a controlled network environment for classifying network traces within monitored time windows. The reason behind such a decision responds to the simplicity of the labeling process.

 
However, the method fails in capturing many of the behavioral details of realistic network traffic. Consequently, the resulting labeled dataset ends up providing a dataset representing partially the conditions observed in a real network environment. Recently, some other methods based on statistical learning, visualization, and a combination of both (assisted methods) have emerged to deal with more realistic network traffic and speed up the labeling process. Nowadays, it is not clear whether such approaches provide a significant help for the labeling process. The fact is that much of the analysis and labeling of network traffic is still performed manually: with an expert user observing the network traces \cite{Huang2020, DIAZVERDEJO202067}. As mentioned by \cite{Catania2012, Sommer2010}, such a situation could be a definite obstacle for the massive adoption of SNIDS in the network security field.

The present document provides an extensive review of the works presenting methodological strategies for generating accurate and representative labels for network security datasets. The survey emphasizes the application of labeling methods based on machine learning and visualization techniques and their benefits and limitations in the generation of quality labels for building and evaluating the performance of SNIDS.

The rest of this document is organized as follows: 
Section \ref{sec:selection_criterial} presents the methodology used for the selection criteria of the papers presented in this survey.
Section \ref{sec:background} provides background information about the labeling process, including a taxonomy and a brief description of the limitations of current labeled datasets available for security research. Then, in Section \ref{sec:labeling-papers}, the current methods for labeling network traffic are reviewed and compared based on the taxonomy, while most relevant aspects of each strategy are discussed in section \ref{sec:discusion}. Section \ref{sec:main_issues} remarks the challenges and open issues in current labeling methods for achieving quality network traffic datasets. Finally, concluding remarks are provided in Section \ref{sec:conclusions}.

\section{Methodology}
\label{sec:selection_criterial}

This study conducts a review of the different labeling methods for generating network traffic datasets and investigates the published journal article in the last 20 years. We performed this  review in two phases. Phase-1 identifies the information resource (search engine) and keywords to execute a query to obtain an initial list of articles. In a second phase, the initial list is filtered under specific selection criteria, and the most related and core articles are stored into the final list reviewed in this article. The purpose of this review article is to answer the following questions: 1) What are the methodologies used by the community to obtain labeled network traffic data sets? 2) What are the recent trends for network traffic labeling methods? 3) What are the characteristics of the established traffic connection labels? 4) What are the benefits and drawbacks of each labeling methodology adopted? 5) What is the future scope of research for creating labeled network traffic datasets?

The queries include terms related to capturing and labeling network traffic datasets and were the result of the authors' experience, as well as the terminology used in prominent literature in this area \cite{Catania2012, Sommer2010}. The terms used during the first stage of the methodology are \textit{traffic dataset}, \textit{dataset labeling}, \textit{intrusion detection}, \textit{network classification}, \textit{dataset creation}, and \textit{labeled dataset}. All the queries aimed at being descriptive for including the creation of network traffic datasets and labeling methodology. Whenever a combination of keywords from both categories was found in the text, the corresponding item was selected as a possible candidate.

The selected queries were applied on the all the best known scientific databases: Scopus \cite{Scopus}, Google Scholar \cite{Scholar}, IEEE Explorer \cite{IEEE}, ACM Digital Library \cite{ACM}, Microsoft Academic Search \cite{MAS}, Springer \cite{Springer} and Mendeley \cite{Mendeley}.  In addition, the proceedings of some of the most important conferences and journals in the field (DEFCON \cite{DEFCON}, USENIX \cite{USENIX}, IPOM \cite{IPOM}, CCS \cite{CCS}, Computers \& Security \cite{C&S}, VizSec \cite{VIS}, EUROSYS \cite{EUROSYS}, and others) were specifically analyzed and also a full-text keyword search was applied. 

\begin{figure}[!h]
\centering
  \includegraphics[width=0.80\linewidth]{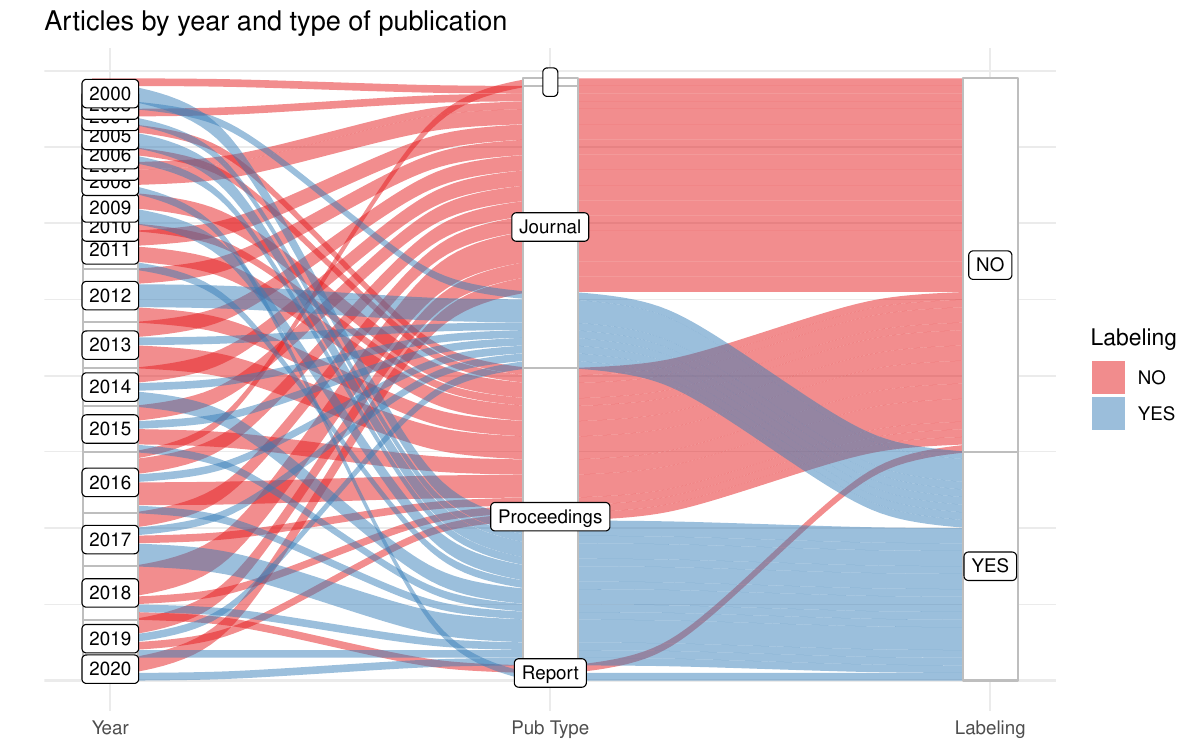}
 \caption{Publication year and type about the selected articles during phase-1.  Articles focused on the methodologies for labeling network traffic datasets are highlighted in blue. }
 \label{fig:stats-selected-papers}
\end{figure}

The first stage of the methodology results in a list of 100 candidate articles. The initial selection criteria for articles were intentionally made with weak constraints so as not to exclude relevant articles and to create a large candidate set. Because of these weak constraints, the initial list contained many false positives that did not meet the predefined criteria. Therefore, in the second stage, the initial list of reviewed and filtered according to i) the generation or capture of labeled network traffic datasets and ii) the methodology for network traffic labeling. To sum up, for an article to be included in this study, it must describe a methodology or framework for getting labeled data related to network traffic traces. On the other hand, those works focused on the creation of unlabeled network traffic datasets or not explicitly stating the labeling methods are omitted from this review. At the end of the second stage, a final list with less than 30 articles was obtained. Basic information (including publication year and type ) about the selected articles during phase-1 and the resulting articles after applying a more strict exclusion criteria is shown in Figure~\ref{fig:stats-selected-papers}. Despite the short number of articles found on the generation of labeled datasets on network traffic, more articles are expected to be published in the future. Especially, given the current interest in the development of machine learning approaches in several other fields.

\section{Background}
\label{sec:background}
Labeling consists of adding one or more meaningful and informative tags to provide context to data \cite{Bernard2017}. 
In the last years, quality dataset labeling has emerged as a fundamental aspect in the application of machine learning models in several areas. The network security field has been focused in the development of NIDS based on machine learning (referred as SNIDS) with the promising goal of achieving better performance detection \cite{Catania2012,Sommer2010}. Consequently, the community has focused in the generation of labeled datasets for analyzing different machine learning approaches in the building of SNIDS. 

This section provides a brief description of SNIDS and the need of quality labeled datasets. Followed by a brief discussion about  the limitations in the current most relevant labeled datasets used for network security. The section ends with a taxonomy for strategies used to create labeled traffic network datasets. 


\subsection{Statistically based NIDS}
\label{sec:snids}
\begin{figure}[!h]
\centering
  \includegraphics[width=0.80\linewidth]{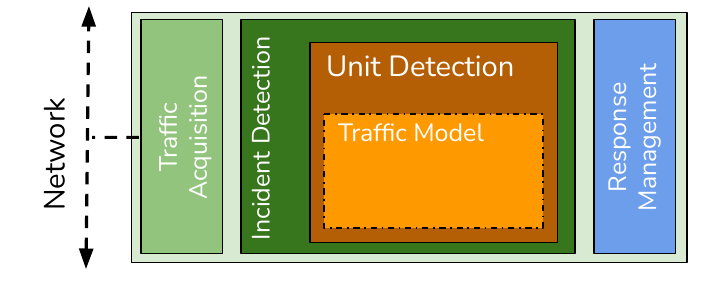}
 \caption{A simplified NIDS architecture (adapted from \cite{Catania2012})}
 \label{fig:SNIDS}
\end{figure}

A simplified NIDS architecture is shown in Figure \ref{fig:SNIDS}. In the first stage, the traffic data acquisition module continuously monitors the traffic, gathers all the network traces on the wire. Then such traces are evaluated by the Incident detector module based on knowledge provided by some predefined Traffic Model. When an incident is detected, an alert is raised, and the suspicious network traces together with information related to the incident are sent to the Response Management module for further expert analysis.

The traditional approach for building a traffic model consists of including a set of rules describing malicious behavior \cite{Roesch1999,Paxson1999}. One of the major inconvenience of rule-based approaches is that rules are capable to recognize only known attacks. Another issue is that rules must be regularly updated by the security experts \cite{Catania2012}.
An alternative approach consists of using a statistical learning model. Under such approach, different network traffic behavior (Email checking, regular social network interaction, Botnet command \& control channel, SPAM, etc.) can be represented as a quantitative response $Y$ while information extracted from network traces such as IP addresses, destination ports, or the number of TCP connections in the last minutes (just to mention a few) are referred as the $p$ different predictors, $X_1$, $X_1$ ... $X_p$. Statistical models assume there is some relationship between $Y$ and $X_{p}$, which can be written in the very general form:

\begin{equation}\label{eq:stat-learning}
Y = f (X) + \epsilon
\end{equation}

Where $f$ is an unknown function of $X_{p}$ that can potentially recognize different network traffic behaviors. In essence, statistical learning refers to a set of approaches for estimating $f$.  One of the most successful methods for estimating $f$ is the so-called supervised learning, where for each observation of the $X$  measurement(s) there is an associated response measurement for $Y$. However,  the performance of statistical learning models following a supervised method will depend on the accuracy and representativeness of the label $Y$.

The inclusion of statistical and machine learning techniques into a NIDS eliminates the need to manually create rules describing traffic behavior by automatically building them from some reference data \cite{Lee1998}.  Another major benefit provided by these methods consists of being able to detect not only known attacks but also their variations. On the other hand,  a major inconvenience with statistically-based NIDS is they require a large amount of labeled network traces to build a traffic model. The difficulty associated with the labeling process of network traffic datasets is one considerable obstacle in the widespread adoption of SNIDS \cite{Catania2012, Sommer2010}.

\subsection{Limitations of current Network traffic Datasets}

When choosing a network traffic dataset to train or test a SNIDS it is necessary to consider the representativeness and accuracy of the network events included. However, obtaining representative and correctly labeled network traffic data sets could be very challenging. Moreover, maintaining such data sets could be prohibitive. The fact is that those organizations capable of producing and publishing representative and accurate data are not very comfortable about the risk of potentially exposing sensitive information. On the other hand, any effort to anonymize data is often considered prohibitively expensive.


For more than 20 years since the first DARPA dataset was published in 1998 \cite{Lippmann2000} there have been numerous published datasets for network security. In a recent survey about labeled datasets for network security, Kenyon et al. \cite{Kenyon2020} enumerates several common flaws in the major public labeled datasets. With more than 27 datasets surveyed,  the lack of labels accuracy and network representation emerge as the most required properties of a high-quality dataset. However, some authors  \cite{ugarte2019} also observe that most of the labeled datasets available for research represent the network behavior for a particular period. Given the rate of change in malicious behaviors,  and the challenge to create and maintain, these labeled datasets become quickly obsolete.  The previously described situation difficult for statistically-based NIDS to generalize its performance to not previously observed attacks. Therefore, more than having only a limited number of high-quality but static labeled datasets, the focus must be on an accurate labeling methodology capable of continuously generating a representative dataset based on network traffic. 

\subsection{A Taxonomy for Labeling network Security Datasets}
\label{sec:labeling-taxonomy}

\begin{figure}[!h]
\centering
  \includegraphics[width=0.95\linewidth]{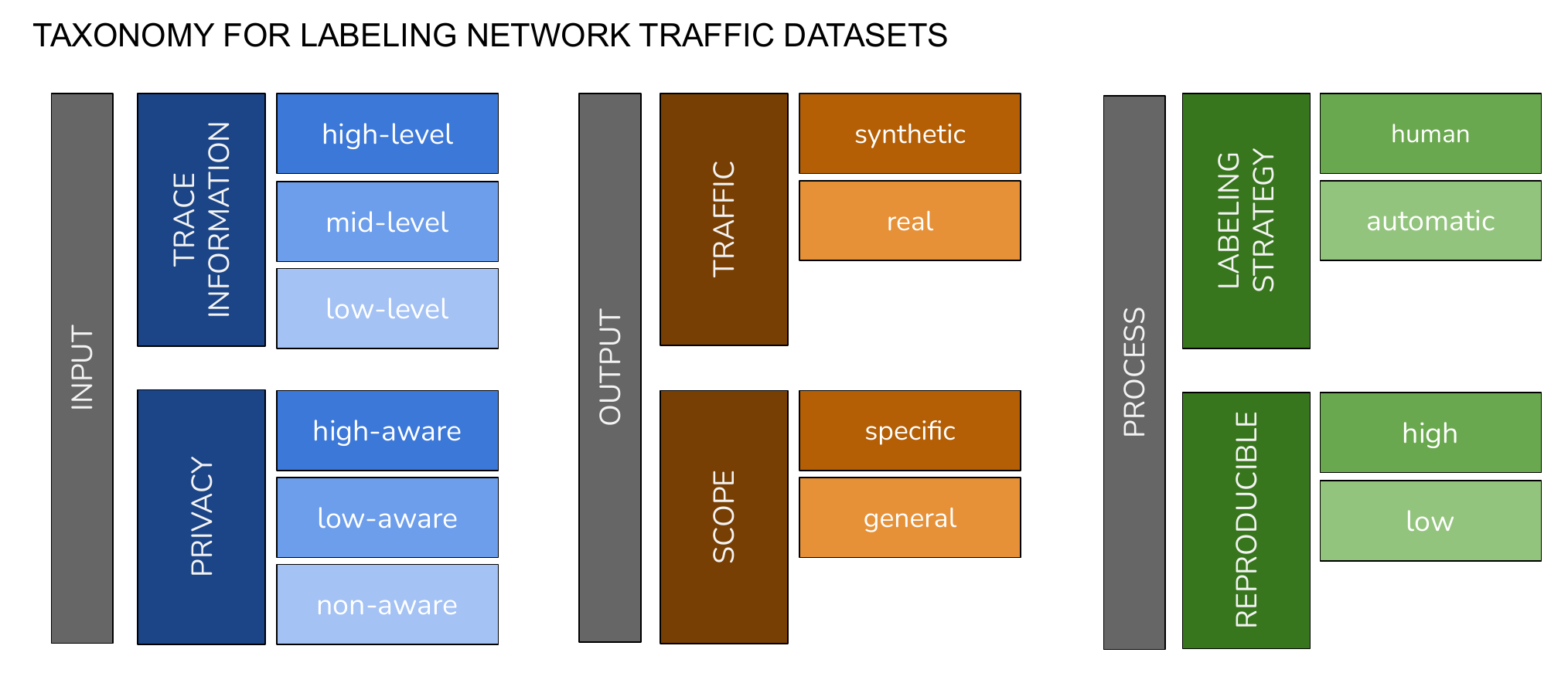}
 \caption{Proposed taxonomy for labeling Network traffic datasets}
 \label{fig:taxonomy}
\end{figure}

Six fundamental categories are considered in the analysis of the labeling methodologies for network security datasets. Figure ~\ref{fig:taxonomy} provides an overview of the six categories and a classification according to three aspects.  i) The data resources used as input, ii) The characteristics of the resulting labeled dataset, and iii) the tools and techniques involved during the labeling process. 

\begin{description}
\item \textbf{Traffic:} the network traffic in the labeled dataset can be categorized into real or synthetic data. The latter refers to data captured from real networks while the former to data artificially generated with the goal of capturing different network conditions.

\item \textbf{Scope:} network traffic datasets can be categorized as \textit{specific-scope:} when the labeling approach focus on a particular network behavior including both normal and malicious (e.g. Garcia et al.~\cite{Garcia2014} aims at capturing only botnet behavior) or as \textit{general-scope:} when no particular consideration has been made during the labeling of the traffic data.

\item \textbf{Trace information}
A  labeling method can be designed for working at different levels on a network trace. At \textit{low-level}, labeling focuses on information directly extracted from a network trace, such as a  list of IP addresses, ports, and network connections, among others. At a \textit{mid level},  the labeling is carried out on network flows level (i.e. labels are per-flow). Finally, at a high-level, labeling is conducted on aggregated information considering interactions or relations between different IP addresses, network flow,  user-level applications, etc.  The methodology and tools involved in the labeling process will depend on the trace information level under analysis. For instance,  high-level trace information such as IP address interactions can be represented as a graph structure. The labeling of such complex structures can require a more elaborated analysis than linear structures such as a list of IP addresses.

\item \textbf{Privacy}
Privacy preservation is another fundamental aspect to consider during the labeling process~\cite{Papadogiannaki2021}. Since dealing with anonymized or encrypted data implies losing potentially valuable information, a labeling methodology will need to be prepared for working under such circumstances. Three categories are considered: \textit{high-privacy-aware},  the labeling methodology is capable of dealing with encrypted or anonymized data. \textit{low-privacy-aware}, the labeling is conducted on the network data not containing the payload. However, sensitive user information such as IP addresses and the port destination remains available, and the last category is
 \textit{non-privacy-aware} when all the network trace information is available during the labeling process.

\item \textbf{Labeling strategy:} Two types of labeling methods are considered:
\textit{human guided labeling} based on human interaction and \textit{automatic labeling} that uses controlled traffic environments. Human guided labeling includes the so-called \textit{manual labeling} which relies only on human expertise (i.e., traditional network traffic analysis with the aid of simple visual charts), and \textit{assisted labeling} which use interactive applications (i.e., a model for recommending labels along with interactive visualizations) Among the three strategies, automatic labeling is the most widely accepted. The general idea behind automatic labeling is to set up a controlled network environment and use the knowledge about the environment to label the traffic.

\item \textbf{Reproducibility} 
A \textit{low} reproducible methodology is designed for being applied only once and producing a unique data set. A \textit{highly} reproducible labeling methodology aims at giving the possibility to a different research team to extend the resulting dataset. For supporting reproducibility,  a labeling methodology should provide detailed information about the tools and resources used during the labeling process.

\end{description}

\section{Current Methods for Labeling Network Traffic}
\label{sec:labeling-papers}
The present section analyzes all the articles collected based on the methodology described in Section \ref{sec:selection_criterial}. The reviewed articles are organized according to the three labeling methods described in section \ref{sec:labeling-taxonomy}.  A summary table with a systematic analysis is presented at the beginning of each section. Each table provides details about the all aspects mentioned in the taxonomy. Then, for each piece of research, a brief description of the labeling approach is provided with a particular focus on the tools and strategies used for conducting the labeling.


\subsection{Automatic Labeling}
\label{sec:automatic-labeling}

\begin{figure}[!h]
\centering
  \includegraphics[width=0.90\linewidth, trim={0 8cm 0 0},clip]{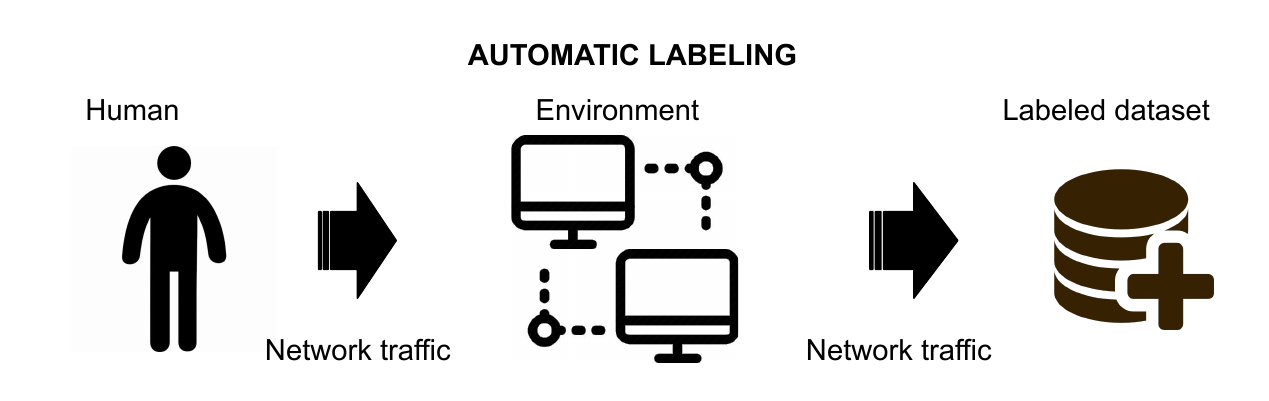} 
 \caption{Under automatic labeling methods labels are the result of monitoring a controlled environment (network infrastructure) by a human (user). By having precise control of the environment, the labeling process can be systematized. }
 \label{fig:automatic-labeling}
\end{figure}

In general, under an automatic labeling technique, the creation of a data set in a controlled and deterministic network environment facilitates recognizing anomalous activities from normal traffic, thus eliminating the process of manual labeling by experts (see Figure \ref{fig:automatic-labeling})

In the last years, several network security researchers have embraced automatic labeling strategies with the help of techniques based on Injection Timing (IT), simulated Behavioral Profiles (BP), and commonly used Network Security tools (NST). Table \ref{tab:automatic-labeling-approches} summarises the surveyed articles using automatic labeling techniques. In addition to the particular labeling technique, the table also provides relevant information about the remaining aspects mentioned in the taxonomy of section \ref{sec:labeling-taxonomy}. 

\begin{table}[t]
\centering
\caption{Summary of  the methodologies using an Automatic strategy for labeling  network traffic. Columns four to eight refer to Reproducibility (\textbf{Prepr.}), Scope, Traffic Type (\textbf{Traffic}), Privacy Awareness \textbf{(Privacy)} and Traces Information \textbf{(Trace)}, as was discussed in the taxonomy.}
\label{tab:automatic-labeling-approches}
\resizebox{\textwidth}{!}{%
\begin{tabular}{@{}llp{3.3cm}lllll@{}}
\midrule
\multicolumn{8}{c}{\cellcolor[HTML]{EFEFEF}\textbf{AUTOMATIC LABELING}}               \\ \midrule
\rowcolor[HTML]{FFFFFF} 
\textbf{Author} & \textbf{Year} & \textbf{ Technique}            & \textbf{Repr.} & \textbf{Scope} & \textbf{Traffic} & \textbf{Privacy} & \textbf{Trace} \\ \midrule

Garcia~\cite{Garcia2014}    & 2014 & IT       & low  & specific & synthetic & non  & mid  \\
Bhuyan~\cite{Bhuyan2015}    & 2015 & IT       & low  & general  & real      & non  & low  \\
Moustafa~\cite{,Moustafa2015}  & 2015 & IT + NST & low  & general  & synthetic & non  & low  \\
Haider~\cite{Haider2017}    & 2017 & IT + NST & low  & general  & synthetic & non  & low  \\
Mukkavilli~\cite{Mukkavilli2016} & 2016 & IT+ BP + NST  & low & general & synthetic  & non & low \\
Lemay~\cite{Lemay2016}     & 2016 & IT       & low  & specific & synthetic & non  & low  \\

Sharafaldin~\cite{Sharafaldin2018}     & 2018  & BP + IT       & low            & general        & synthetic        & low              & low                \\
Shiravi~\cite{Shiravi2012}         & 2012  & BP + IT       & low            & general        & synthetic        & low              & low                \\
Lippmann~\cite{Lippmann2000}  & 2000 & BP + NST  & low  & general  & synthetic & non  & low  \\
Stolfo~\cite{Stolfo2000}    & 2000 & BP + NST         & low  & general  & synthetic & non  & high \\
Catania~\cite{Catania2012b}   & 2012 & NST      & low  & general  & real      & non  & low  \\
Gargiulo~\cite{Gargiulo2012}  & 2012 & NST      & high & general  & real      & low  & low  \\
Song~\cite{Song2011}      & 2011 & NST                    & low  & general  & real*     & non  & low  \\
Sperotto~\cite{Sperotto2009}  & 2009 & NST                    & low  & general  & real      & low  & low  \\
A-Navarro~\cite{Aparicio-Navarro2014} & 2014 & NST      & low  & general  & real      & non  & low  \\
Ring~\cite{Ring2017361}      & 2017 & NST + BP         & low  & general  & synthetic & high & mid  \\
Sangster~\cite{Sangster2009}  & 2009 & NST + BP         & low  & general  & synthetic & non  & low  \\ \midrule
\end{tabular}
}
\end{table}

\subsubsection{Injection Timing}

One of the most successful methods for labeling network traffic datasets consists of generating different network traces at specific time windows. Then label all the network traces accordingly to the specific target behavior.  This technique is known as Injection Timing (IT)~\cite{Lemay2016}.  

\begin{figure}[!h]
\centering
  \includegraphics[width=0.8\linewidth]{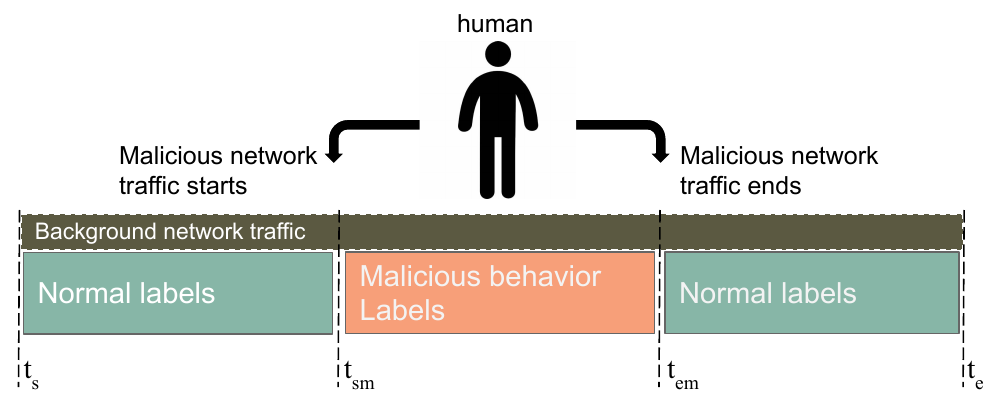}
 \caption{An example of the Injection Timing labeling strategy using specific time windows for injecting and labeling malicious network traffic.}
 \label{fig:injection_timing}
\end{figure}

The injection timing strategy requires a controlled network environment where the user has precise information about the different applications generating traffic on the network.   Figure(Figure \ref{fig:injection_timing}) provides a simplified overview of the injection timing strategy. At $t_s$ the network has just become operational for the first time. At the time $t_{sm}$ the user injects into the network particular malware traffic (DDOS, Botnet, port scanning, etc.) Since the network has just become operational. All the background traffic from the time window from $t_s$ to $t_{sm}$ is labeled as normal. Beyond $t_{sm}$ all network traffic is labeled as malicious. When the user stops injecting the malicious behavior at $t_{em}$, all the traffic becomes labeled as normal again until the network is permanently shut down at time $t_e$.

Since the Injection Timing strategy is applied under controlled environments, it is possible to create labeled datasets with numerous network traffic behaviors. This feature provides some level of authenticity for validating the experimental results on the generated labeled dataset. However, since labels are obtained by merely contrasting the execution time window of each generated network trace, a strict time control mechanism is necessary for obtaining accurate labels.

A clear example of the application of the injection timing strategy is observed in the work of Garcia et al. \cite{Garcia2014}. In particular, the authors use Injection Timing for generating a labeled dataset focused on Botnet attacks (The CTU-13 Dataset).  A topology consisting of a set of virtualized computers with the Microsoft Windows XP SP2 operating system on a Linux Debian host was used for capturing the traffic through time windows. 

For the particular case of Botnet network behavior, with Injection timing, it is easier to label these network traces with higher accuracy than other types of attacks. The fact is that Botnets tend to have a temporary and very localized behavior, which means that most actions remain unchanged for several minutes. Therefore, the separation of the traffic into time windows facilitates the control of the botnet behavior of the network.

Following a similar approach, Bhuyan et al. \cite{Bhuyan2015} use a subset of the TUIDS (Tezpur University Intrusion Detection System) testbed network for capturing normal and malicious traffic traces. Normal traffic is collected independently from real users of the networks. At the same time, several types of malicious network traffic are injected by infecting several network stations. All the network traces from these stations are then captured considering the specific time intervals. Then after a pre-processing to extract traffic from those infected stations, all network traces are mixed using the time interval windows of each connection to classify them as either \textit{malicious} or \textit{non-malicious}.

Tools like IXIA PerfectStorm are also used for generating labeled data.  By using this tool, it is possible to generate up to 9 families of malware. The Australian Defence Force Academy relay on the IXIA PerfectStorm tool for its labeling strategy.  The captured traffic from IXIA perfectStorm and label the traces using the windows time interval and the attack information reported by the tool. This strategy was used for generating the UNSW-NB15 (Moustafa et al. \cite{Moustafa2015}) and NGIDS-DS ( \cite{Haider2017}) datasets. 

On the other hand, Mukkavilli et al. \cite{Mukkavilli2016} focus on labeling traffic interaction between users and cloud services (i.e. Amazon EC2). The authors rely on the PlanetLab infrastructure. PlanetLab is a (now-extint) group of computers available as a testbed for computer networks and distributed systems \cite{PlanetLab,planetlab2006}. The authors use specific PlanetLab nodes mimicking normal users and their interaction with cloud services based on normal and uniform traffic distributions. All traffic from normal nodes is labeled as normal. On the other hand,  malicious behavior is injected at different time windows to generate malicious traffic. In particular,  DDOS, port scanning, and ARP spoofing attacks take place at random intervals, while normal behavior traffic continues to run continuously.  All attack nodes have the precaution to start the malicious behavior within the same time window with a minimum delay and labeled accordingly. Several other authors follow a similar approach combining tools for simulating normal behavior with injection timing techniques for capturing malicious~\cite{Ring2017361,Shiravi-Survey2012,Sharafaldin2018}.


The work of Lemay et al. \cite{Lemay2016} applies an Injection Timing strategy for labeling the traffic from SCADA (Supervisory Control and Data Acquisition) systems. Due to the sensitive nature of these networks, there was little publicly available data. Through the use of pre-established and simulated network architecture, the authors could generate Modbus \cite{modbus2008} network traffic with precise knowledge about the behavior observed on each network trace type. Then if a packet is part of a trace group that includes malicious activity, it will be labeled as 'malicious.' Otherwise, it is labeled as 'normal.'

\subsubsection{Behavioral Profiles}
The use of Behavioral Profiles is another strategy used for automatically labeling network traffic. Behavioral profiles provide the information to simulate a specific feature or aspect of the network. A profile encompasses an abstract representation of features and events of real-world behaviors considered from the network perspective \cite{Shiravi2012}. Therefore, profiles are usually implemented as computer programs executing common tasks according to some previously defined mathematical model (usually a probability distribution). These profiles are then used by human agents or operators to simulate the specific events in the network. Their abstract property effectively makes them network-agnostic and allows them to be applied to different setup and topologies. Thus, the labeling process using this technique is straightforward; all the traffic generated by a profile simulating normal traffic will be labeled as normal. Similarly, all the traffic generated from a malicious behavioral profile is labeled as malicious. 

In this way, Shiravi et al. \cite{Shiravi2012} combine two classes of profiles to generate a labeled dataset with different characteristics and events. A first profile \textit{A} tries to describe an attack scenario unambiguously. While a second profile - \textit{B} - encapsulates distributions and mathematical behaviors extracted from certain entities and represented as procedures with pre and postconditions, thus representing normal traffic. Examples include the statistical distributions of packet sizes of a protocol, number of packets per flow, specific patterns in the payload, size of payload, the request time distribution of protocol.

Sharafaldin et al. \cite{Sharafaldin2018}, also focus on two behavioral profiles. An abstract benign profile is built upon 25 user behaviors based on the HTTP, HTTPS, FTP, SSH, and email protocols. The benign profile is responsible for modeling human interactions' abstract behavior and generating naturalistic benign background traffic. Six malicious profiles are generated based on frequent attacks. Then, by combining these profiles, several labeled datasets can be generated, each one with unique characteristics for the evaluation. By merely altering the combination of the profiles, it is possible to control the composition (e.g., protocols) and statistical characteristics (e.g., request times, packet arrival times, burst rates, volume) of the resulting data set.

Other works \cite{Lippmann2000,Lippmann2000b,Ring2017361,Mukkavilli2016,Sangster2009} combines behavioral profiles with other techniques for improving the representativeness of the resulting labeled datasets. In particular, Lippmann et al. \cite{Lippmann2000, Lippmann2000b} proposes the use of automata to simulate traffic behaviors with a traffic distribution similar to the observed between a small Air Force base and the Internet. Through custom software automata in the test bed, hundreds of programmers, secretaries, managers, and other types of users running common UNIX and Windows NT application programs are simulated. Automata interact with high-level user application programs or they implement clients for network services such as HTTP, SMTP, and POP3. Low-level TCP/IP protocol interactions are handled by kernel software and are not simulated.

\subsubsection{Network Security Tools}
The labeling process is carried out based on the information provided by network security tools (NST) such as sniffers, honeypots, or even using a NIDS.

The application of a NST labeling strategy was one of the strategy applied in the generation of the DARPA datasets (1998-99) \cite{Lippmann2000,Lippmann2000b} and the KDD99 \cite{Stolfo2000}. As part of the DARPA IDS evaluation program, a testbed was created with many types of live traffic using virtual hosts to simulate a small Air Force base separated by a router from the Internet. Different types of attacks were conducted outside the network and captured by a sniffer located in the network router. Any network trace coming outside the network (the internet) is considered as malicious while those from inside are normal.

A similar approach was more recently applied by Ring et al. \cite{Ring2017361} combining a virtualized network environment with traces captured from a real web server exposed to the internet. Normal traffic is generated through behaviors profiles while Malicious traffic comes from the external server as well as particular attacks injected from the virtualized infrastructure. The authors applied anonymization techniques netflow information for guaranteeing privacy.

NST tools are also applied for labeling traffic in capture-the-flag competition. Ideally,  capture-the-flag competitions are a valuable source for gathering distributed normal and malicious traffic.  However, by default,  data sets do not contain labels. Sangster et al.~\cite{Sangster2009} apply a set of pre-established rules and user roles in combination with network sniffers to capture several traffic behaviors. Then, based on that information, they provide the correct label to each network trace. The malicious traffic is captured from specific computers belonging to NSA security experts. On the other hand, Normal traffic is generated artificially using profile behaviors.

Gargiulo et al. \cite{Gargiulo2012} and Catania et al.~\cite{Catania2012b} use rule-based NIDS for generating labels. In particular, Catania et al. analyze the performance of stand-alone NIDS for labeling traffic and provide some results when results labels are used training SNIDS. On the other hand,  Gargiulo et al. use the principles of Dempster-Shafer's theory for combining the information of several rule-based NIDS.  In their approach, the authors use the basic probability assignment for calculating the final decision about a particular network trace. The resulting labels are then used for training a SNIDS.

The work of Navarro et al. \cite{Aparicio-Navarro2014}  propose a similar strategy for automatic labeling 802.11 network traffic using a NIDS based on unsupervised anomalies. The NIDS analyzes the traffic, and for each of the network traces, the system provides three numerical values with information about the label's belief. The belief values represent the probability of observing normal or attack behaviors. A third value is used for registering how uncertain the system is about network label and adjusting the system accordingly. The labeling process is established by using a threshold of possible values per label. The threshold is set using the mean and standard deviation of the probability values set by NIDS. Through this threshold, all network traces whose label value is not within this threshold will be discarded from the dataset due to the degree of belief they present. Connections within the value range will be labeled according to the NIDS's highest probability between the Normal and Malicious classes. Thus, the resulting labeled dataset will contain those connections that NIDS determined with a high degree of confidence.

On the other hand, Sperotto et al. \cite{Sperotto2009} with TWENTE, and Song et al. \cite{Song2011} with KU aim to provide the security community with more realistic data sets. Their labeling method is based on the analysis of several honeypots with different architectures inserted within a network environment. Then, all captured traffic to specific monitored services in the honeypots can be easily labeled as malicious. By using honeynets, there is no human interference during the data collection process (i.e., any form of attack injection is prevented). Therefore, the attacks present in the dataset reflect the situation of a real network.

\subsection {Human-Guided Labeling}

\begin{figure}[!h]
\centering
  \includegraphics[width=0.90\linewidth, trim={0 0 0 8cm},clip]{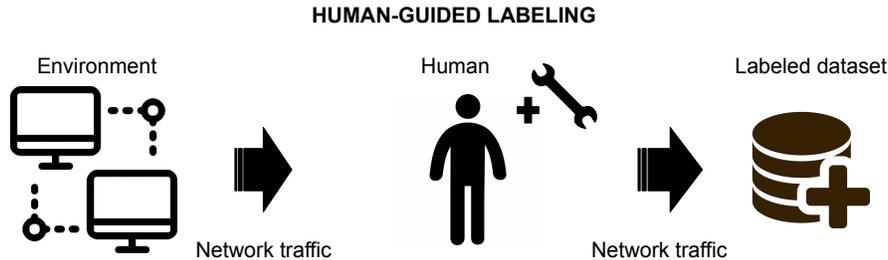} 
 \caption{Under human-guided labeling methods, the environment (network infrastructure) is not controlled by a human (user). Labels are the result of human knowledge with the eventual assistance of particular tools. }
 \label{fig:human-labeling}
\end{figure}

Many authors \cite{aladin, Chifflier2012, Fan2019, Gornitz2013} consider human experience is an essential aspect of traffic analysis and the subsequent connection labeling. Therefore, under human-guided labeling methods, the network environment is not controlled and all the work relies in the expert users. However, since experts are an invaluable resource, labeling time has to be efficiently used (see Figure \ref{fig:human-labeling}).

\subsubsection{Manual}
\label{sec:user_interaction}
A very significant percentage of today's network analysis is performed manually (i.e., without assistance from any system) by security experts \cite{Shiravi-Survey2012}. Manual labeling by network traffic experts requires a precise understanding of the network behavior for differentiating between malicious and normal traces. Unfortunately, many of these extensive network analysis processes are not published, and despite being widely used, the research community has limited knowledge about it.

There are several approaches to reduce human effort in the manual labeling process. Some authors propose the use of visual tools (Viz) to improve traffic behavior analysis. Others suggest collaborative environments between a label prediction model and security experts (Crowd). A summary of the mos relevant approaches are presented in Table~\ref{tab:manual-labeling-approches}.

\begin{table}[t]
\centering
\caption{ Summary of  the methodologies using a manual strategy for labeling network traffic. Columns four to eight refer to Reproducibility (\textbf{Prepr.}), Scope, Traffic Type (\textbf{Traffic}), Privacy Awareness \textbf{(Privacy)} and Traces Information \textbf{(Trace)}, as was discussed in the taxonomy.
}
\label{tab:manual-labeling-approches}
\resizebox{\textwidth}{!}{%
\begin{tabular}{@{}llp{3cm}lllll@{}}
\midrule
\rowcolor[HTML]{EFEFEF} 
\multicolumn{8}{c}{\cellcolor[HTML]{EFEFEF}\textbf{MANUAL LABELING}}  \\ \midrule
\rowcolor[HTML]{FFFFFF} 
\textbf{AUTHOR} & \textbf{YEAR} & \textbf{ TOOL}            & \textbf{REPR.} & \textbf{SCOPE} & \textbf{TRAFFIC} & \textbf{PRIVACY} & \textbf{TRACE} \\ \midrule
                 
Koike~\cite{Koike2006}     & 2006 & Viz                   & low  & general  & real      & low  & low  \\
Livnat~\cite{Livnat2005}    & 2005 & Viz                   & low  & general  & real      & low  & mid  \\
Ren~\cite{Ren2005}       & 2005 & Viz                   & low  & general  & real      & low  & high \\
Scott~\cite{Scott2003}     & 2003 & Viz                   & low  & general  & real      & low  & high \\
Chen~\cite{Chen2014}     & 2014 & Viz + Crowd    & low  & general  & real      & low  & mid \\
Huang~\cite{Huang2020}     & 2020 & Crowd         & low  & general  & real      & non  & mid  \\ \midrule

\end{tabular}
}
\end{table}

Many works try to reduce the effort involved during manual labeling through the use of visualization tools. In particular, the use of visual systems help the user during labeling by improving correlation between malicious patterns and making the user more confident about their labels \cite{Cappers2018}.

NIVA \cite{Scott2003} is one of the first examples of a data visualization tool for intrusion detection. NIVA uses information from various intrusion detectors and incorporates references and colors to give the attacks a significant value. The color of the reference represents the severity of the attacks. Yellow is moderate, while red is the most severe.

On the other hand, IDGraphs \cite{Ren2005}, uses the visualization technique called \textit{Histographs}: a visual technique to map the brightness of a pixel to the frequency of the data. By mapping multiple combinations of features in the input data, attacks with different characteristics can be identified. IDGraphs not only shows an overview of the underlying network data, but also allows an in-depth analysis of possible anomalies through dynamic queries \cite{Shiravi-Survey2012}.

Livnat et al. \cite{Livnat2005} develop a chord-based visualization for the correlation of network alerts. The approach is based on the notion that an alert must possess three attributes: what, when, and where. These attributes can be used as a basis for comparing heterogeneous events. A network topology map is located at the center with the various alert records in a surrounding ring. The ring's width represents time and is divided into several periods of the history of each connection. A line is drawn from an alert type on the outer ring to a particular host on the topology map to represent a triggered alarm. Thicker lines show a more significant number of alerts of a single type, and the larger nodes in the topology map represent hosts that experience unique alerts.

The authors of IPMatrix \cite{Koike2006} believe that an attacker's IP address, even if falsified, is a significant factor in an attack, and administrators can take appropriate countermeasures based on it. Using a combination of heatmap and scatter plots, IP Matrix represents the full range of IPs. IP Matrix incorporates two 256x256 matrices. The first, the\textit{ Internet level} matrix, only maps the first two octets while the \textit{local level} matrix maps the last two octets, allowing the local and Internet level IP addresses to be seen simultaneously. Each alert generated by an IDS is mapped using a pixel within its appropriate cell. Pixels are color-coded to represent attacks of different nature, but because a pixel is too small to be seen, the background of a cell is colored with the most frequent attack type. A disadvantage of this system is that there are no connections between the local level and the Internet hosts, which makes the system less intuitive.

Other manual approaches make use of crowdsourcing (Crowd) for obtaining labeled datasets.  Crowdsourcing has proven to be a cost-effectively way to obtain a large-scale labeled dataset. Moreover, in some fields, it has been demonstrated that nonexpert annotations were relatively useful for training a statistical model~\cite{Zhang2019}.

Chen et al.~\cite{Chen2014} introduce the OCEANS (Online Collaborative Explorative Analysis on Network Security) system, which integrates visual analytics methods and collaboration features as a web application. In particular, OCEANS integrates the crowd input from security experts and makes everyone contribute to labeling the network events. OCEANS visuals offer detail of individual network flows, including IP, port, time, and network attributes from both source and destination side, as well as health status and IPS logs. OCEANS provides a web interface where a user can submit events while others can view and comment on them. Users need to provide label information describing the event. Then all the crowd input is synthesized into an event graph and event timeline. The tool provides a suspicious score based on the number of events an IP address involved, adding the count of agreement on this event and subtracting the count of disagreement.

More recently, Huang et al.~\cite{Huang2020} developed and released IoT inspector. An open-source application for capturing and labeling network traffic from smart home devices.  The application crowdsources the data from within home networks and provides a mechanism for simplifying the labeling. The resulting dataset is not focused on malicious behavior, but on modeling the particular behavior of the different brands of smart devices. Contrary to desktop computers, smart home devices perform very specific tasks making their networking behavior very predictable~\cite{Shahid2018}.

\subsubsection{Assisted}
\label{sec:active_learning}
To facilitate the analysis and subsequent labeling of network traffic by experts, several authors have proposed using a technique called Active Learning (AL) \cite{Yang2015, McElwee2017}. 

\begin{figure}[!h]
\centering
  \includegraphics[width=0.95\linewidth]{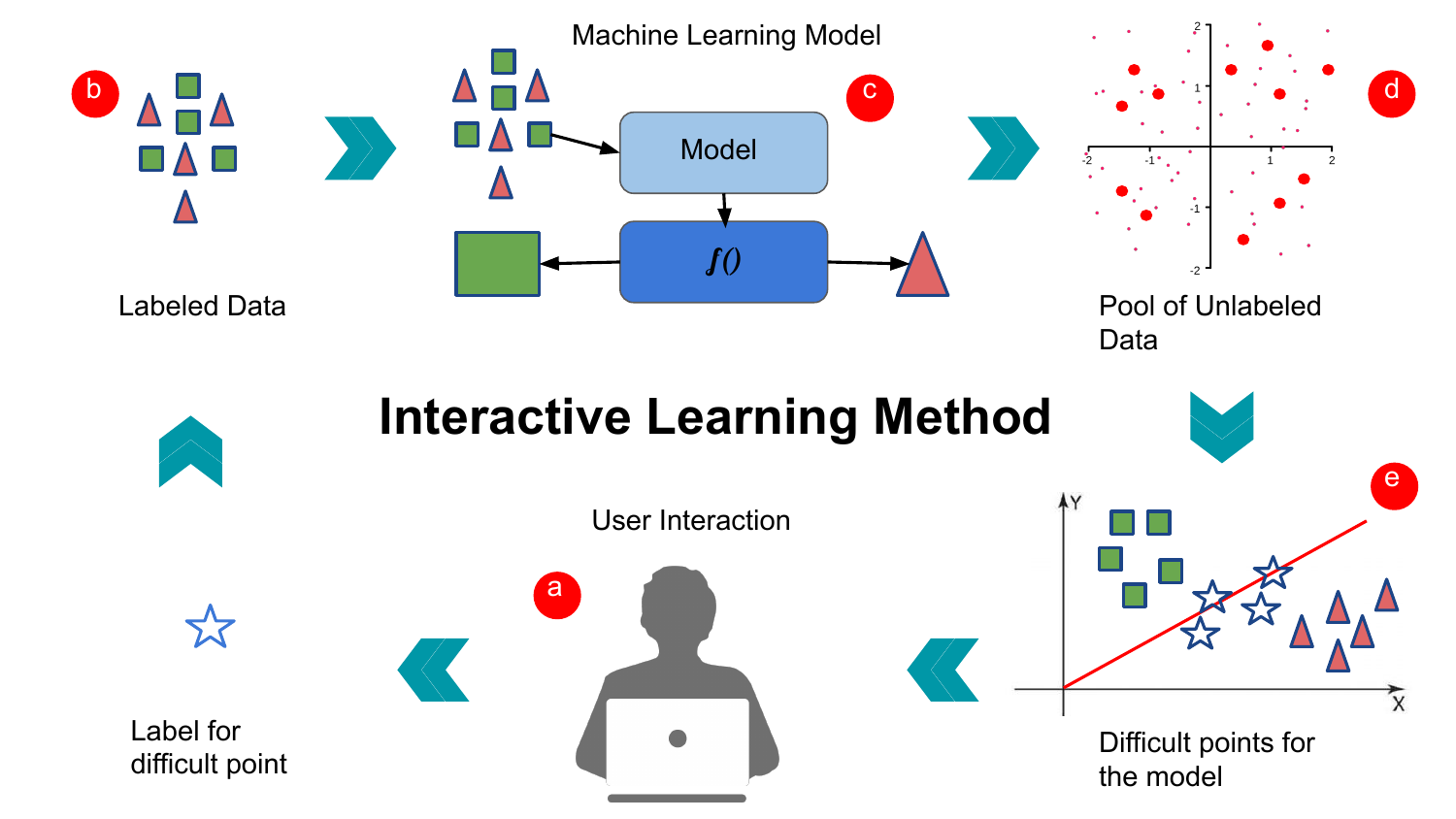}
 \caption{Work cycle of an Active Learning Strategy used for labeling}
 \label{fig:active_learning}
\end{figure}

AL refers to human-in-the-loop methods, where a prediction model is iteratively updated with input from expert users (see Figure \ref{fig:active_learning}). The expert user (a) is responsible for taking decisions on those connections where the model has a high degree of uncertainty (e) (i.e. those connections near the decision boundary) than expected (a strategy known as Uncertainty Sampling \cite{Yang2015, LEWIS1994148}). These final decisions are used for labeling the data (b) fed into the model (c) to improve its objective function and prediction performance on unlabeled data (d). 

A summary of the most relevant approaches is presented in Table~\ref{tab:assisted-labeling-approches}.  In some cases,  AL is combined with other labeling tools. In particular, the combination of Visual (Viz) and AL techniques has emerged as an effective approach for labeling network traffic.

\begin{table}[b!]
\centering
\caption{ Summary of  the methodologies using a assisted strategy for labeling network traffic. Columns four to eight refer to Reproducibility (\textbf{Prepr.}), Scope, Traffic Type (\textbf{Traffic}), Privacy Awareness \textbf{(Privacy)} and Traces Information \textbf{(Trace)}, as was discussed in the taxonomy.
}
\label{tab:assisted-labeling-approches}
\resizebox{\textwidth}{!}{%
\begin{tabular}{@{}llp{3cm}lllll@{}}
\midrule
\rowcolor[HTML]{EFEFEF} 
\multicolumn{8}{c}{\cellcolor[HTML]{EFEFEF}\textbf{ASSISTED LABELING}}                \\ \midrule
\rowcolor[HTML]{FFFFFF} 
\textbf{AUTHOR} & \textbf{YEAR} & \textbf{ TOOL}            & \textbf{REPR.} & \textbf{SCOPE} & \textbf{TRAFFIC} & \textbf{PRIVACY} & \textbf{TRACE} \\ \midrule
Almgren~\cite{Almgren2004}   & 2004 & AL                     & low  & general  & real      & non  & high \\
Beaugnon~\cite{Beaugnon2017}  & 2017 & AL + Viz               & high & general  & real      & low  & high \\
Fan~\cite{Fan2019}       & 2019 & AL + Viz               & low  & specific & real      & low  & mid  \\
Guerra~\cite{guerra2019vizsec}    & 2019 & AL + Viz               & high & specific & real      & high & high \\
Gornitz~\cite{Gornitz2013}   & 2013 & AL                     & low  & specific & real      & non  & low  \\
McElwee~\cite{McElwee2017}   & 2017 & AL                     & low  & general  & real      & low  & high \\
Stokes~\cite{aladin}   & 2008 & AL                     & low  & general  & real      & non  & high \\
\midrule
\end{tabular}%
}
\end{table}

Classical AL techniques are widely used in labeling large volumes of data in general, and it has started to be used for constructing labeled network traffic datasets. 

In their work from 2004, Almgren and Jonsson  \cite{Almgren2004} propose a classical AL strategy based on uncertainty sampling \cite{Yang2015, LEWIS1994148} to select the most suitable network traces to be labeled by the expert users.

On the other hand, other works attempt to accelerate the AL working cycle by including several strategies for improving the quality of the network data to be labeled by expert-users. Stokes \cite{aladin} includes a rare category detection algorithm \cite{Pelleg2004} into to AL work cycle to encourage the discovery of families of network traces sharing the same features. Similarly, Görnitz uses a k-nearest neighbors (KNN) approach to identify various network trace families. Both approaches guarantee that every family has representative members during the expert labeling process and reduces the sampling bias.
Beaugnon et al. \cite{Beaugnon2017} also rely on rare category detection to avoid sampling bias. Moreover, they apply a divide-and-conquer strategy during labeling to ensure good expert-model interaction focused on small traffic sections.

Similarly, McElwee \cite{McElwee2017} proposes an AL intrusion detection method based on Random Forests and k-Means clustering algorithms. The daily events are submitted to a Random Forests classifier and events receiving more than 95\% of the votes are considered correct and saved into a \textit{master} dataset. The remaining events, conforms a candidate dataset grouped into $k$ groups using k-means clustering. Each group is analyzed and classified by an expert and then saved into the master dataset.

Other works combine a visualization component with the AL labeling strategy. The motivation behind including visual components is to improve the user experience during the AL work cycle. A better user experience translates into better quality labels for the prediction model.
Xin Fan et al. \cite{Fan2019} present one of the most recent approaches combining AL techniques with a visual tool to provide the user with a better representation of the traffic being analyzed. The authors use a graph to display a two-dimensional topological representation of the network connections. The nodes in the graph are differentiated by color to identify the connection type quickly and a color intensity matrix to show the interaction between the connections. Several other visual tools such as histograms and boxplots are employed during the labeling process. Histograms are used for representing the percentage of the traffic of the various protocols/ports. Boxplots are used to show the distributions of the destination ports and the number of records of the different IPs 

In the work of Beaugnon et al. \cite{Beaugnon2017,Chifflier2012}, the authors also implements a visual representation for the user interaction process. In this case, the visual application provides a mechanism for organizing the network traffic in different groups. A set of queries and filters facilitates the user to create families of connections for further analysis by small network traffic groups.

Otherwise, Guerra et al. present RiskID \cite{guerra2019vizsec, Guerra2019}, a modern application focus in the labeling of real traffic. Specifically, RiskID intend to create labeled datasets based in botnet and normal behaviors. The RiskID application uses visualizations to graphically encode features of network connections and promote visual comparison. A visualization display whole traffic using a heatmap representation based in features. The heatmap promotes the search of pattern inside the traffic with similar behaviors. Other visualization shows statistical report for a punctual connection using color-map, histogram and a pie-chart. In the background, two algorithms are used to actively organize connections and predict potential labels: a recommendation algorithm and a semi-supervised learning strategy (AL strategy). These algorithms together with interactive adaptions to the user interface constitute a behavior recommendation.


\section{Discussion}
\label{sec:discusion}

No matter the labeling strategy used, they focused on the accuracy and representativeness of the resulting datasets. However, despite their frequent use, there are still substantial problems inherent to the labeling methodologies.  Significant aspects such as privacy, reproducibility, and the level of expertise required are not discussed in depth during the implementation of each strategy. Table \ref{tab:comparing-labeling-approaches} summarizes the more significant aspects of the three labeling strategies.

\begin{table}[t]
\centering
\caption{Benefits and drawbacks of the strategies for labeling network traffic datasets.}
\label{tab:comparing-labeling-approaches}
\resizebox{\textwidth}{!}{%
\begin{tabular}{@{}lll@{}}
\\
\toprule
\textbf{Labeling Strategy} & \textbf{Benefits} & \textbf{Drawbacks} \\ \midrule
\cellcolor[HTML]{EFEFEF}\textbf{Automatic} &
  \begin{tabular}[c]{@{}l@{}} 
  \tabitem Very fast \\ 
  \tabitem Easy to adapt to new specific behaviors \\ 
  \tabitem Low expertise \\
  \tabitem Moderate accuracy \\ 
  \end{tabular} &
  \begin{tabular}[c]{@{}l@{}}

  \tabitem Low representativeness \\ 
  \tabitem Hard to reproduce \\ 
  \tabitem Non Privacy\end{tabular} \\
                           &               &               \\
\cellcolor[HTML]{EFEFEF}\textbf{Manual} &
  \begin{tabular}[c]{@{}l@{}}
  \tabitem High representiveness\\ 
  \tabitem High Accuracy
  \end{tabular} &
  \begin{tabular}[c]{@{}l@{}}
  \tabitem Slow\\ 
  \tabitem High expertise\\ 
  \tabitem Hard to reproduce\\ 
  \tabitem Low Privacy\end{tabular} \\
                           &               &               \\
\cellcolor[HTML]{EFEFEF}\textbf{Assisted} &
  \begin{tabular}[c]{@{}l@{}}
  \tabitem Fast\\ 
  \tabitem Medium expertise\\ 
  \tabitem High representivenes\\ 
  \tabitem Moderate Accuracy\end{tabular} &
  \begin{tabular}[c]{@{}l@{}}
  \tabitem Hard to reproduce\\ 
  \tabitem Hard to adapt to new specific behaviors\\ 
  \tabitem Low Privacy\end{tabular} \\ \bottomrule
\end{tabular}%
}
\end{table}

\subsection{Automatic Labeling}


Automatic labeling strategies are the preferred approach to obtain labeled network traffic data. Such a decision responds to the low level of expertise required and the relative speed for generating large volumes of labeled network traffic. The fact is that automatic labeling strategies do not require a high level of expertise compared to manual labeling techniques. 

Among all the automatic labeling strategies, the Injection Timing strategy is the simplest and straightforward. Unfortunately, this strategy shows several limitations regarding the critical representativeness required in the data. The main limitation is that malicious and normal traffic activities were usually captured from two different and uncorrelated environments. When both captures are merged and collectively analyzed, it could be easy to discriminate malicious from normal traffic. The background traffic, routing information, and the hosts present in the network are some aspects to be considered when capturing network traffic from several sources. Another significant issue with injection timing is the lack of a clear approach for supporting privacy awareness. Although, it is theoretically possible to apply several anonymization techniques, the fact is that most of the articles implemented the techniques has not even considered  a methodology for protecting the privacy in background and normal traffic  ( \cite{Garcia2014, Lemay2016, Bhuyan2015}). The only exception is the work of Ring et al.~\cite{Ring2017361} where the authors have discussed IP anoymization techniques during the labeling process.


On the other hand, the labeling process based on Network security tools is usually applied on real traffic, which provides a better representiveness. However, it could be difficult to ensure the required accuracy. As is the case of the work of Navarro et al. \cite{Aparicio-Navarro2014}, and Gargiulo et al. \cite{Gargiulo2012} who use a NIDS based on a set of rules for describing malicious behavior.  Both authors use only those connections classified by NIDS with a high confidence rank. These approaches guarantee the reliability of the labels in the resulting dataset but neglect those connections that are difficult to predict and that are very useful to improve detection systems. To mitigate this bias towards easy-to-detect connections, Navarro proposes the use of an expert for manually analyzing and labeling just those connections with a high degree of hesitation.

In general, all reviewed labeling approaches based on NIDS  \cite{Stolfo2000, Gargiulo2012, Aparicio-Navarro2014, Haider2017,Catania2012b} rely on some ruleset that needs to be periodically updated. i.e., Whenever a new variant of a malicious behavior emerges, an expert needs to write a new rule describing such behavior. The fact is that there is no guarantee the traffic generating an alert in NIDS do not contain an attack. Therefore, those supposedly normal traffic traces should be analyzed in depth before added to the final labeled dataset.

The honeynets alternatives \cite{Song2011, Sperotto2009,Ring2017361} provide a straightforward procedure for labeling malicious network traces. However, similarly to the NIDS approaches, it shows serious flaws for labeling normal traffic. The simple rule of considering all traffic captured from the honeypots as malicious \cite{Sperotto2009} does not guarantee the rest of the traffic is free of undetected malicious behaviors.

The fact is that ensuring the quality of the automatic labeling methods remains a challenging task. In Lemay et al. \cite{Lemay2016}, they consider that if a packet is part of a connection including malicious activity, it has to be labeled as malicious. Otherwise, it is labeled as normal. However, when an attacker connects to an FTP service for sending an \textit{exploit}, not all the traffic contains malicious behavior. Under a deeper inspection of packet capture, it can be argued that the TCP connection needed to connect to the service to send the \textit{exploit} is not malicious. After all, the connection procedure is no different from other legitimate connections established by other clients to the server. In that case, only the packets containing the actual \textit{exploit} should be labeled as malicious. A similar problem can be found in Bhuyan et al. \cite{Bhuyan2015}, where the authors attempt to generate normal traffic with varied characteristics from traffic captures of users' daily activities. Malicious traffic is generated by launching attacks and infecting different users' servers. Under this scenario, it is not easy to guarantee that all the traffic captured comes from users is normal. The fact is that considering that the network is clean before the first attack occurs is a mere assumption. 
To sum up, Automatic Labeling methods provide a fast and simple approach for generating a considerable amount of labeled traffic. They can easily adapt to new behavior without a high level of expertise. However, the deployment of the infrastructure for capturing and labeling traffic could be difficult to reproduce. Moreover, it is clear that despite all the precautions during the generation of synthetic traffic, these methods still have serious drawbacks regarding the level of representativeness and label accuracy. Ideally, a network traffic labeled dataset should not exhibit any inconsistent property of the network infrastructure and its relying traffic. The traffic must look as realistic as possible, including both normal and malicious traffic behaviors. In particular, traffic data should be free of noise and not include any leakage caused by the selected labeling strategy. Therefore, the Automatic Labeling method should implement a detailed specification of the capture processes to provide coherent and valuable traffic data.



\subsection{Human-guided labeling}
\label{sec:human-guided-labeling}
In general, the manual labeling methods generate datasets with good representativeness and accuracy. The main inconvenience relies on the difficulty of labeling the traffic volume required for current SNIDS needs. Users with high expertise are a fundamental resource during the labeling process.
Recent approaches including visualization techniques and interactive labeling methods have emerged to facilitate the incorporation of users with a lower degree of expertise. 

However, those manual labeling approaches relying only on visualization suffer from the same drawback ~\cite{Koike2006, Livnat2005, Ren2005, Scott2003}. They still require a high level of expertise for performing the actual classification. Despite having attracted considerable attention for identifying malicious activities \cite{Shiravi-Survey2012}, their adoption in real-world applications has been hampered by their complexity. 


Human-guided methods based on AL strategies aim at improving the speed of the labeling process while keeping high representativeness of the resulting data. The inclusion of a statistical learning model can be a valuable tool for helping the user during the decision process. Moreover, some of the approaches~\cite{Fan2019,guerra2019vizsec} claim the expertise required for using such systems is reduced. 
Nevertheless, the role of the expert remains a fundamental aspect for guaranteeing the quality of the labels. The expert is responsible for generating the initial set of labels required for training the prediction model. Moreover, the expert is responsible for labeling during the AL working cycle when a connection is difficult to discriminate between normal or malicious. The precision of these labels could impact the overall accuracy of the recommendations made by the relatively simple models based on Logistic Regression \cite{Chifflier2012}, Fuzzy c-means algorithm \cite{Fan2019} or Random Forests \cite{McElwee2017}.  

Manual labeling strategies are difficult to reproduce and extend to new traffic behaviors. In many cases, these strategies will rely only on the ability of the expertise of the user to recognize new behaviors. Similar is the case of assisted approaches, although to a minor degree. In some cases, if the distribution of the new traffic behavior significantly differs from the known distribution, the prediction model has to be retrained to recognize new behaviors. Moreover, when very focused visualization techniques are combined with AL, adapting them to new traffic behavior could not be straightforward.
On the other hand, privacy awareness under the surveyed manual approaches remains under minimal standards. None of them discuss the consequences of traffic encryption or anonymization during the labeling. However, in many cases, the labeling is conducted through observing mid-level trace information such as net flows~\cite{Livnat2005,Fan2019,Huang2020,Chen2014}, which indicates that payload information is not available during labeling.  Similarly, complex visuals such as ~\cite{Ren2005,Scott2003}  are suitable for hiding considerable private information and still being useful for labeling. 
Not differently is the case of assisted labeling strategies, where most of them seem not specially prepared for dealing with privacy mechanisms. Only the work of Guerra et al.~\cite{guerra2019vizsec} have considered the inclusion of anonymized network traces during the labeling process. 

To sum up, all the human-guide labeling methods seem to be more well-suited for label network traffic with high accuracy and representativeness. However, despite the considerable improvements, these strategies still show several issues regarding the capacity for rapid and continuous labeling of network traffic.

\section{Open Issues}
\label{sec:main_issues}

\subsection{Deficiencies in the representativeness of labeling strategies}

Since DARPA \cite{Lippmann2000, Lippmann2000b}, there have been several attempts to improve the quality of network traffic labeled datasets. However, there are still several problems regarding the representativeness of the network scenarios. The fact is that automatic labeling strategies have serious issues for operating on real traffic ~\cite{Garcia2014, Lemay2016, Bhuyan2015, Shiravi2012, Sharafaldin2018}. Even those strategies using NST such as honeynets which capture real attacks suffer from representativeness problems when they try to incorporate normal traces into the resulting labeled dataset. Therefore, these strategies cannot represent all the details about traffic dynamics and potential real-world network attacks. As shown in \cite{Hofstede2018}, the network traffic differs between lab environments and production networks.

On the other hand, human-guided labeling strategies certainly improve the authenticity of the resulting labels. However, the labeling process is still slow and challenging for obtaining a sufficient number of representative labels for use on current SNIDS implementation. 

Privacy preservation is another major issue regarding human-guide labeling strategies. In manual and assisted strategies, the expert has access to all the traffic information from real users. The previous situation is not so critical in automatic labeling strategies, since normal traffic is usually generated artificially~\cite{Sharafaldin2018,Shiravi2012} or under controlled conditions~\cite{Garcia2014}.
 
A partial solution consisted of applying anonymization techniques during the capture process. Therefore, network traffic can be subjected to encryption or attribute extraction procedures for hiding different portions of the traffic during the labeling process. \cite{Cermak2018}. Many human-guided strategies rely on this approach for a minimal privacy preservation.  Almgren and Fan \cite{Almgren2004, Fan2019} for instance, perform traffic labeling at the flow level, hiding relevant information such included in the payload. Similarly, Beaugnon et al. \cite{Beaugnon2017} perform a complete per-flow feature extraction procedure depriving the community of using the entire network payload. However, the main problem with anonymization is that the removal of valuable information from the network has an impact on the correct representation of network behavior. When dealing with real and representative labeled datasets generation, it is essential to ensure precise and consistent network traffic information. The process requires careful monitoring and capturing of the different aspects of regular traffic, in conjunction with a fast and accurate labeling method for providing the SNIDS and the research community with an adequate dataset.

It seems that the inclusion of collaborative approaches~\cite{Chen2014} is an aspect that could improve human-guided labeling techniques.  Firstly, the incorporation of multiple users in the labeling is a significant improvement of the overall speed of the process. Alleviating one of the main drawbacks of human-guided strategies. Secondly, it incorporates into the process more evaluative analysis on the different behaviors, allowing both differentiation and unification of criteria. In this way, through a kind of voting, labels could be established with greater accuracy while keeping representativeness at a high level.  On the other hand, the speed of collaborative labeling techniques impacts AL-based strategies. Since traces are labeled faster, the prediction model gets earlier feedback, which accelerates the phases of the learning cycle. 

Another possible path for improving the speed and quality of human-guided strategies is the use of users' labeling history. It would be straightforward for current human-guided strategies~\cite{Chifflier2012, Beaugnon2017, Guerra2017, Fan2019}  to create a matrix of user preferences in relation to the set of traces that make up the traffic. The resulting matrix can be used to build a Recommendation System to focus the labeling on groups of traces with similar characteristics and according to the user's preferences. In this way, the whole labeling process is enhanced.



\subsection{Support for systematic periodical updates}

Recently some members of the network security community have started to mention that due to the evolution of malicious behavior and the constant innovations in attack strategies, network traffic labeled datasets need to be updated periodically \cite{Nehinbe2011, Kenyon2020}.  However, from all labeling strategies in section ~\ref{sec:labeling-papers}, only a few of them provide a consistent approach to continuously updating dataset information and preventing it from expiring over time. 

The automatic labeling strategies require the deployment of a complex network infrastructure. The maintenance of such infrastructures complicates the extension to new behaviors. Moreover,  the whole reproducibility of the process is adversely affected, since infrastructure, user profiles, the malware used are usually not precisely described, 

On the other hand, some assisted labeling strategies seem to be more adaptable to new behaviors, as they depend on the generalization of their prediction model. Both Beaugnon et al. ~\cite{Chifflier2012} and Guerra et al.~\cite{Guerra2019} published the source code of their visualization tools together with the AL prediction model. However, the model performance can often decay since predictions are biased to specific network behavior. Updating these models require a continuous execution of the AL working cycle, which demands expert user assistance.

Consistently with the previous section, a collaborative approach can also be applied to guarantee a certain degree of reproducibility during the expert interaction. In the best case, if a network trace received different classifications, but most of them are from a particular behavior, it can be estimated as the correct behavior and finally set the label.

In general, given the  volume, velocity and variety  characteristics of network traffic~\cite{Wang2021},  it is necessary to move away from strategies that result in static datasets. Having a continuous pipeline for generating accurate and representative labeled datasets is part of the so-called MLOps (Machine Learning Operations). MLOps \cite{banerjee2020challenges}, is a recent field of machine learning that aims to make building and deploying models more systematic. Current labeling strategies need to incorporate MLOps strategies capable of adapting to current traffic distributions and intrusions approaches and provide modifiable, extensible, and reproducible mechanisms for continuous labeled dataset delivery.

\subsection{Lack of consistent validation methodologies}
Despite the strategy employed for labeling a dataset, a consistent methodology is necessary to validate its results. The components of this methodology should be adapted depending on the applied strategy.

In the methods based on automatic labeling, the most common evaluation methodology is based on the similarity against real traffic. Several authors \cite{Sharafaldin2018, Gharib2016} proposed similarity metrics for evaluating the resulting datasets. Metrics such as complete network configuration, labeling accuracy, available protocols, attack diversity, and metadata provide a quality standard for a dataset. However, the impact of the labeled dataset quality in creating network behavior classification models remains unknown. 

In contrast, the validation of methods based on human-guided labeling is considerably more complex. It is necessary to evaluate the components included in the work cycle and the interaction between them to determine the effectiveness of the strategy. Unfortunately, AL strategies discussed in the \ref{sec:discusion} section do not analyze the benefits and the problems involved in the work cycle of labeling data. Surveyed articles ~\cite{aladin, Chifflier2012, Fan2019, McElwee2017}  do not include any process for measuring the accuracy of the prediction model as the AL cycle progresses. Other important considerations, such as the minimum number of labels needed to make accurate suggestions or how the strategy reacts when noisy data is introduced, are not explored in depth.

Similarly is the case for those strategies including visualization tools. The main goal behind these strategies is to assist the user during the labeling process. However, most of the reviewed works considering visualization tools \cite{Ren2005, Livnat2005, Scott2003, Koike2006, Chifflier2012, Fan2019} have not evaluated the benefits and usefulness of the proposed visualizations. Fan et al. \cite{Fan2019}, and Guerra et al. \cite{guerra2019vizsec} are among the few authors to analyze the performance of different visualization techniques used to improve pattern perception during the interactive process. The fact is that the availability and cost of conducting a validation with expert users and traffic analysts affect the evaluation process.  As a result, analytical and empirical evaluations of the systems often do not provide the information needed to establish the usefulness of the support tools. 

It seems critical that the community starts to focus on providing user studies to measure the impact of the tools in the labeling process and get relevant information about the labeling strategy followed by users. Such studies should include information about the expertise level of the users, their interaction with the assistant tools, and the human effort associated with the complete labeling process. 

Finally, current labeling strategies must provide an in-depth analysis of the correlation between labeling strategy, label quality, and the final performance of the resulting detection models

\section{Conclusions}
\label{sec:conclusions}
Labeled dataset generation is a fundamental resource for network security research. However, all current labeling strategies experience significant problems in terms of quality, volume, and speed. There is a trade-off between the quality of the resulting labeled dataset and the amount of network traces included. Automatic labeling method provide a large amount of labeled network traces, but the accuracy and representative could not be guaranteed. Human-guided method are an improvement for the quality of resulting labeled dataset, but since they still heavily depend on user expertise, the speed and volume of labeled data could be insufficient.

A more significant problem is that the current methodologies are oriented to create a static version of the datasets. A static labeled dataset is only suitable for research during a very short time period. The development of a validated methodology including a continuous pipeline for incorporating new representative and accurate network traces is fundamental for continue with the development of network security research. In the case of Statistically-based NIDS, the need of a standard strategy for a continuous generation of quality labeled datasets is entirely accordant with the recent MLOps roles included in the production cycle beyond the network security field.

To sum up, quality labeled datasets are not enough. The network security research community need to standardize the methodology reducing expert-user interaction with focus on reproducible and continuous validation in concordance of the data-centric models used nowadays when deploying machine learning products in real-life scenarios.



\section{Acknowledgements}

The authors would like to thank the financial support received by Argentinean ANPCyT- FONCYT through the project PICT 1435-2015 and the Argentinean National Scientific and Technical Research Council.

\bibliographystyle{elsarticle-num-names}
\bibliography{jguerra-survey.bib}

\begin{thebibliography}{91}
\expandafter\ifx\csname natexlab\endcsname\relax\def\natexlab#1{#1}\fi
\providecommand{\url}[1]{\texttt{#1}}
\providecommand{\href}[2]{#2}
\providecommand{\path}[1]{#1}
\providecommand{\DOIprefix}{doi:}
\providecommand{\ArXivprefix}{arXiv:}
\providecommand{\URLprefix}{URL: }
\providecommand{\Pubmedprefix}{pmid:}
\providecommand{\doi}[1]{\href{http://dx.doi.org/#1}{\path{#1}}}
\providecommand{\Pubmed}[1]{\href{pmid:#1}{\path{#1}}}
\providecommand{\bibinfo}[2]{#2}
\ifx\xfnm\relax \def\xfnm[#1]{\unskip,\space#1}\fi
\bibitem[{Resende and Drummond(2018)}]{Resende2018}
\bibinfo{author}{P.~A.~A. Resende}, \bibinfo{author}{A.~C. Drummond},
\newblock \bibinfo{title}{{A survey of random forest based methods for
  intrusion detection systems}},
\newblock \bibinfo{journal}{ACM Computing Surveys} \bibinfo{volume}{51}
  (\bibinfo{year}{2018}). \DOIprefix\doi{10.1145/3178582}.
\bibitem[{Glass-Vanderlan et~al.(2018)Glass-Vanderlan, Iannacone, Vincent,
  Chen, and Bridges}]{Glass-Vanderlan2018}
\bibinfo{author}{T.~R. Glass-Vanderlan}, \bibinfo{author}{M.~D. Iannacone},
  \bibinfo{author}{M.~S. Vincent}, \bibinfo{author}{Q.~Chen},
  \bibinfo{author}{R.~A. Bridges},
\newblock \bibinfo{title}{{A survey of intrusion detection systems leveraging
  host data}},
\newblock \bibinfo{journal}{arXiv} \bibinfo{volume}{52} (\bibinfo{year}{2018}).
  \href{http://arxiv.org/abs/1805.06070}{{\tt arXiv:1805.06070}}.
\bibitem[{Buczak and Guven(2016)}]{Buczak2016}
\bibinfo{author}{A.~L. Buczak}, \bibinfo{author}{E.~Guven},
\newblock \bibinfo{title}{{A Survey of Data Mining and Machine Learning Methods
  for Cyber Security Intrusion Detection}},
\newblock \bibinfo{journal}{IEEE Communications Surveys and Tutorials}
  \bibinfo{volume}{18} (\bibinfo{year}{2016}) \bibinfo{pages}{1153--1176}.
  \DOIprefix\doi{10.1109/COMST.2015.2494502}.
\bibitem[{Catania and {Garcia Garino}(2012)}]{Catania2012}
\bibinfo{author}{C.~Catania}, \bibinfo{author}{C.~{Garcia Garino}},
\newblock \bibinfo{title}{{Automatic network intrusion detection: Current
  techniques and open issues}},
\newblock \bibinfo{journal}{Computer and Electrical Engineering}
  \bibinfo{volume}{7} (\bibinfo{year}{2012}) \bibinfo{pages}{1063 -- 1073}.
\bibitem[{Vasilomanolakis et~al.(2015)Vasilomanolakis, Karuppayah, Muhlhauser,
  and Fischer}]{Vasilomanolakis2015}
\bibinfo{author}{E.~Vasilomanolakis}, \bibinfo{author}{S.~Karuppayah},
  \bibinfo{author}{M.~Muhlhauser}, \bibinfo{author}{M.~Fischer},
\newblock \bibinfo{title}{{Taxonomy and survey of collaborative intrusion
  detection}},
\newblock \bibinfo{journal}{ACM Computing Surveys} \bibinfo{volume}{47}
  (\bibinfo{year}{2015}) \bibinfo{pages}{1--33}.
  \DOIprefix\doi{10.1145/2716260}.
\bibitem[{Sharafaldin et~al.(2018)Sharafaldin, {Habibi Lashkari}, and
  Ghorbani}]{Sharafaldin2018}
\bibinfo{author}{I.~Sharafaldin}, \bibinfo{author}{A.~{Habibi Lashkari}},
  \bibinfo{author}{A.~A. Ghorbani},
\newblock \bibinfo{title}{{Toward Generating a New Intrusion Detection Dataset
  and Intrusion Traffic Characterization}},
\newblock \bibinfo{journal}{International Conference on Information Systems
  Security and Privacy}  (\bibinfo{year}{2018}) \bibinfo{pages}{108--116}.
  \DOIprefix\doi{10.5220/0006639801080116}.
\bibitem[{Maciá-Fernández et~al.(2018)Maciá-Fernández, Camacho,
  Magán-Carrión, García-Teodoro, and Therón}]{MaciaFernandez2018}
\bibinfo{author}{G.~Maciá-Fernández}, \bibinfo{author}{J.~Camacho},
  \bibinfo{author}{R.~Magán-Carrión}, \bibinfo{author}{P.~García-Teodoro},
  \bibinfo{author}{R.~Therón},
\newblock \bibinfo{title}{Ugr‘16: A new dataset for the evaluation of
  cyclostationarity-based network idss},
\newblock \bibinfo{journal}{Computers \& Security} \bibinfo{volume}{73}
  (\bibinfo{year}{2018}) \bibinfo{pages}{411 -- 424}.
  \DOIprefix\doi{https://doi.org/10.1016/j.cose.2017.11.004}.
\bibitem[{Hofstede et~al.(2014)Hofstede, Čeleda, Trammell, Drago, Sadre,
  Sperotto, and Pras}]{Hofstede}
\bibinfo{author}{R.~Hofstede}, \bibinfo{author}{P.~Čeleda},
  \bibinfo{author}{B.~Trammell}, \bibinfo{author}{I.~Drago},
  \bibinfo{author}{R.~Sadre}, \bibinfo{author}{A.~Sperotto},
  \bibinfo{author}{A.~Pras},
\newblock \bibinfo{title}{Flow monitoring explained: From packet capture to
  data analysis with netflow and ipfix},
\newblock \bibinfo{journal}{IEEE Communications Surveys Tutorials}
  \bibinfo{volume}{16} (\bibinfo{year}{2014}) \bibinfo{pages}{2037--2064}.
  \DOIprefix\doi{10.1109/COMST.2014.2321898}.
\bibitem[{Kumar and Dutta(2016)}]{Kumar}
\bibinfo{author}{S.~Kumar}, \bibinfo{author}{K.~Dutta},
\newblock \bibinfo{title}{Intrusion detection in mobile ad hoc networks:
  techniques, systems, and future challenges},
\newblock \bibinfo{journal}{Security and Communication Networks}
  \bibinfo{volume}{9} (\bibinfo{year}{2016}) \bibinfo{pages}{2484--2556}.
  \URLprefix \url{https://onlinelibrary.wiley.com/doi/abs/10.1002/sec.1484}.
  \DOIprefix\doi{https://doi.org/10.1002/sec.1484}.
\bibitem[{Sun et~al.(2007)Sun, Osborne, Xiao, and Guizani}]{Sun}
\bibinfo{author}{B.~Sun}, \bibinfo{author}{L.~Osborne},
  \bibinfo{author}{Y.~Xiao}, \bibinfo{author}{S.~Guizani},
\newblock \bibinfo{title}{Intrusion detection techniques in mobile ad hoc and
  wireless sensor networks},
\newblock \bibinfo{journal}{IEEE Wireless Communications} \bibinfo{volume}{14}
  (\bibinfo{year}{2007}) \bibinfo{pages}{56--63}.
  \DOIprefix\doi{10.1109/MWC.2007.4396943}.
\bibitem[{Zarpelão et~al.(2017)Zarpelão, Miani, Kawakani, and {de
  Alvarenga}}]{ZARPELAO201725}
\bibinfo{author}{B.~B. Zarpelão}, \bibinfo{author}{R.~S. Miani},
  \bibinfo{author}{C.~T. Kawakani}, \bibinfo{author}{S.~C. {de Alvarenga}},
\newblock \bibinfo{title}{A survey of intrusion detection in internet of
  things},
\newblock \bibinfo{journal}{Journal of Network and Computer Applications}
  \bibinfo{volume}{84} (\bibinfo{year}{2017}) \bibinfo{pages}{25--37}.
  \DOIprefix\doi{https://doi.org/10.1016/j.jnca.2017.02.009}.
\bibitem[{Pham et~al.(2016)Pham, Hoang, and Van~Canh}]{Pham}
\bibinfo{author}{T.~S. Pham}, \bibinfo{author}{T.~H. Hoang},
  \bibinfo{author}{V.~Van~Canh},
\newblock \bibinfo{title}{Machine learning techniques for web intrusion
  detection — a comparison},
\newblock in: \bibinfo{booktitle}{2016 Eighth International Conference on
  Knowledge and Systems Engineering (KSE)}, \bibinfo{year}{2016}, pp.
  \bibinfo{pages}{291--297}. \DOIprefix\doi{10.1109/KSE.2016.7758069}.
\bibitem[{da~Costa et~al.(2017)da~Costa, Barbon, Miani, Rodrigues, and
  Zarpelão}]{Costa}
\bibinfo{author}{V.~G.~T. da~Costa}, \bibinfo{author}{S.~Barbon},
  \bibinfo{author}{R.~S. Miani}, \bibinfo{author}{J.~J. P.~C. Rodrigues},
  \bibinfo{author}{B.~B. Zarpelão},
\newblock \bibinfo{title}{Detecting mobile botnets through machine learning and
  system calls analysis},
\newblock in: \bibinfo{booktitle}{2017 IEEE International Conference on
  Communications (ICC)}, \bibinfo{year}{2017}, pp. \bibinfo{pages}{1--6}.
  \DOIprefix\doi{10.1109/ICC.2017.7997390}.
\bibitem[{Tesfahun and {Lalitha Bhaskari}(2013)}]{Tesfahun2013}
\bibinfo{author}{A.~Tesfahun}, \bibinfo{author}{D.~{Lalitha Bhaskari}},
\newblock \bibinfo{title}{{Intrusion detection using random forests classifier
  with SMOTE and feature reduction}},
\newblock \bibinfo{journal}{Proceedings - 2013 International Conference on
  Cloud and Ubiquitous Computing and Emerging Technologies, CUBE 2013}
  (\bibinfo{year}{2013}) \bibinfo{pages}{127--132}.
  \DOIprefix\doi{10.1109/CUBE.2013.31}.
\bibitem[{Yueai and Junjie(2009)}]{Yueai}
\bibinfo{author}{Z.~Yueai}, \bibinfo{author}{C.~Junjie},
\newblock \bibinfo{title}{Application of unbalanced data approach to network
  intrusion detection},
\newblock in: \bibinfo{booktitle}{2009 First International Workshop on Database
  Technology and Applications}, \bibinfo{year}{2009}, pp.
  \bibinfo{pages}{140--143}. \DOIprefix\doi{10.1109/DBTA.2009.116}.
\bibitem[{Wheelus et~al.(2014)Wheelus, Khoshgoftaar, Zuech, and
  Najafabadi}]{Wheelus}
\bibinfo{author}{C.~Wheelus}, \bibinfo{author}{T.~M. Khoshgoftaar},
  \bibinfo{author}{R.~Zuech}, \bibinfo{author}{M.~M. Najafabadi},
\newblock \bibinfo{title}{A session based approach for aggregating network
  traffic data -- the santa dataset},
\newblock in: \bibinfo{booktitle}{2014 IEEE International Conference on
  Bioinformatics and Bioengineering}, \bibinfo{year}{2014}, pp.
  \bibinfo{pages}{369--378}. \DOIprefix\doi{10.1109/BIBE.2014.72}.
\bibitem[{Haddadi and Zincir-Heywood(2016)}]{Haddadi}
\bibinfo{author}{F.~Haddadi}, \bibinfo{author}{A.~N. Zincir-Heywood},
\newblock \bibinfo{title}{Benchmarking the effect of flow exporters and
  protocol filters on botnet traffic classification},
\newblock \bibinfo{journal}{IEEE Systems Journal} \bibinfo{volume}{10}
  (\bibinfo{year}{2016}) \bibinfo{pages}{1390--1401}.
  \DOIprefix\doi{10.1109/JSYST.2014.2364743}.
\bibitem[{McKeown et~al.(2008)McKeown, Anderson, Balakrishnan, Parulkar,
  Peterson, Rexford, Shenker, and Turner}]{McKeown}
\bibinfo{author}{N.~McKeown}, \bibinfo{author}{T.~Anderson},
  \bibinfo{author}{H.~Balakrishnan}, \bibinfo{author}{G.~Parulkar},
  \bibinfo{author}{L.~Peterson}, \bibinfo{author}{J.~Rexford},
  \bibinfo{author}{S.~Shenker}, \bibinfo{author}{J.~Turner},
\newblock \bibinfo{title}{Openflow: Enabling innovation in campus networks},
\newblock \bibinfo{journal}{SIGCOMM Comput. Commun. Rev.} \bibinfo{volume}{38}
  (\bibinfo{year}{2008}) \bibinfo{pages}{69–74}. \URLprefix
  \url{https://doi.org/10.1145/1355734.1355746}.
  \DOIprefix\doi{10.1145/1355734.1355746}.
\bibitem[{Cugola and Margara(2012)}]{Cugola2012}
\bibinfo{author}{G.~Cugola}, \bibinfo{author}{A.~Margara},
\newblock \bibinfo{title}{{Processing flows of information: From data stream to
  complex event processing}},
\newblock \bibinfo{journal}{ACM Computing Surveys} \bibinfo{volume}{44}
  (\bibinfo{year}{2012}). \DOIprefix\doi{10.1145/2187671.2187677}.
\bibitem[{Huang et~al.(2020)Huang, Apthorpe, Li, Acar, and
  Feamster}]{Huang2020}
\bibinfo{author}{D.~Y. Huang}, \bibinfo{author}{N.~Apthorpe},
  \bibinfo{author}{F.~Li}, \bibinfo{author}{G.~Acar},
  \bibinfo{author}{N.~Feamster},
\newblock \bibinfo{title}{Iot inspector: Crowdsourcing labeled network traffic
  from smart home devices at scale},
\newblock \bibinfo{journal}{Proceedings of the ACM on Interactive, Mobile,
  Wearable and Ubiquitous Technologies} \bibinfo{volume}{4}
  (\bibinfo{year}{2020}). \DOIprefix\doi{10.1145/3397333}.
\bibitem[{Díaz-Verdejo et~al.(2020)Díaz-Verdejo, Estepa, Estepa,
  Madinabeitia, and Muñoz-Calle}]{DIAZVERDEJO202067}
\bibinfo{author}{J.~E. Díaz-Verdejo}, \bibinfo{author}{A.~Estepa},
  \bibinfo{author}{R.~Estepa}, \bibinfo{author}{G.~Madinabeitia},
  \bibinfo{author}{F.~J. Muñoz-Calle},
\newblock \bibinfo{title}{A methodology for conducting efficient sanitization
  of http training datasets},
\newblock \bibinfo{journal}{Future Generation Computer Systems}
  \bibinfo{volume}{109} (\bibinfo{year}{2020}) \bibinfo{pages}{67--82}.
  \DOIprefix\doi{https://doi.org/10.1016/j.future.2020.03.033}.
\bibitem[{Sommer and Paxson(2010)}]{Sommer2010}
\bibinfo{author}{R.~Sommer}, \bibinfo{author}{V.~Paxson},
\newblock \bibinfo{title}{{Outside the Closed World: On Using Machine Learning
  for Network Intrusion Detection}},
\newblock \bibinfo{journal}{IEEE Symposium on Security and Privacy}
  \bibinfo{volume}{0} (\bibinfo{year}{2010}) \bibinfo{pages}{305--316}.
  \DOIprefix\doi{10.1109/SP.2010.25}.
\bibitem[{B.V(????)}]{Scopus}
\bibinfo{author}{E.~B.V}, \bibinfo{title}{Scopus},
  \bibinfo{howpublished}{\url{https://www.scopus.com/}}, ????
  \bibinfo{note}{[Online; accessed July-2019]}.
\bibitem[{Google(????)}]{Scholar}
\bibinfo{author}{Google}, \bibinfo{title}{Google scholar},
  \bibinfo{howpublished}{\url{https://scholar.google.com/}}, ????
  \bibinfo{note}{[Online; accessed July-2019]}.
\bibitem[{IEEE(????)}]{IEEE}
\bibinfo{author}{IEEE}, \bibinfo{title}{Ieee explorer, advancing technology of
  humanity}, \bibinfo{howpublished}{\url{https://www.ieee.org/}}, ????
  \bibinfo{note}{[Online; accessed July-2019]}.
\bibitem[{for Computing~Machinery(????)}]{ACM}
\bibinfo{author}{A.~for Computing~Machinery}, \bibinfo{title}{Acm digital
  library}, \bibinfo{howpublished}{\url{https://dl.acm.org/}}, ????
  \bibinfo{note}{[Online; accessed July-2019]}.
\bibitem[{Microsoft(????)}]{MAS}
\bibinfo{author}{Microsoft}, \bibinfo{title}{Microsoft academic search},
  \bibinfo{howpublished}{\url{https://academic.microsoft.com/home}}, ????
  \bibinfo{note}{[Online; accessed July-2019]}.
\bibitem[{Nature(????)}]{Springer}
\bibinfo{author}{S.~Nature}, \bibinfo{title}{Springer},
  \bibinfo{howpublished}{\url{https://www.springer.com/}}, ????
  \bibinfo{note}{[Online; accessed July-2019]}.
\bibitem[{B.V(????)}]{Mendeley}
\bibinfo{author}{E.~B.V}, \bibinfo{title}{Mendeley brings your research to
  life, so you can make an impact on tomorrow},
  \bibinfo{howpublished}{\url{https://www.mendeley.com/}}, ????
  \bibinfo{note}{[Online; accessed July-2019]}.
\bibitem[{DEF(1993)}]{DEFCON}
\bibinfo{title}{Def conference},
  \bibinfo{howpublished}{https://www.defcon.org/}, \bibinfo{year}{1993}.
  \bibinfo{note}{[Online; accessed May-2021]}.
\bibitem[{USENIX(1975)}]{USENIX}
\bibinfo{author}{USENIX}, \bibinfo{title}{The advanced computing systems
  association}, \bibinfo{howpublished}{https://www.usenix.org/},
  \bibinfo{year}{1975}. \bibinfo{note}{[Online; accessed May-2021]}.
\bibitem[{Workshop(2009)}]{IPOM}
\bibinfo{author}{I.~I. Workshop}, \bibinfo{title}{Ipom: International workshop
  on ip operations and management},
  \bibinfo{howpublished}{https://link.springer.com/book/10.1007/978-3-642-04968-2},
  \bibinfo{year}{2009}. \bibinfo{note}{[Online; accessed May-2021]}.
\bibitem[{for Computing~Machinery(1993)}]{CCS}
\bibinfo{author}{A.~A. for Computing~Machinery}, \bibinfo{title}{Computer and
  communications security},
  \bibinfo{howpublished}{https://dl.acm.org/conference/ccs},
  \bibinfo{year}{1993}. \bibinfo{note}{[Online; accessed May-2021]}.
\bibitem[{ELSEVIER(2000)}]{C&S}
\bibinfo{author}{ELSEVIER}, \bibinfo{title}{Computers \& security. the
  international source of innovation for the information security and it audit
  professional},
  \bibinfo{howpublished}{https://www.journals.elsevier.com/computers-and-security},
  \bibinfo{year}{2000}. \bibinfo{note}{[Online; accessed May-2021]}.
\bibitem[{IEEE(2000)}]{VIS}
\bibinfo{author}{IEEE}, \bibinfo{title}{Ieee vis: Visualization \& visual
  analytics}, \bibinfo{howpublished}{http://ieeevis.org/},
  \bibinfo{year}{2000}. \bibinfo{note}{[Online; accessed May-2021]}.
\bibitem[{for Computing~Machinery(2006)}]{EUROSYS}
\bibinfo{author}{A.~A. for Computing~Machinery}, \bibinfo{title}{European
  conference on computer systems},
  \bibinfo{howpublished}{https://dl.acm.org/conference/eurosys},
  \bibinfo{year}{2006}. \bibinfo{note}{[Online; accessed May-2021]}.
\bibitem[{Bernard et~al.(2018)Bernard, Hutter, Zeppelzauer, Fellner, and
  Sedlmair}]{Bernard2017}
\bibinfo{author}{J.~Bernard}, \bibinfo{author}{M.~Hutter},
  \bibinfo{author}{M.~Zeppelzauer}, \bibinfo{author}{D.~Fellner},
  \bibinfo{author}{M.~Sedlmair},
\newblock \bibinfo{title}{Comparing visual-interactive labeling with active
  learning: An experimental study},
\newblock \bibinfo{journal}{IEEE Transactions on Visualization and Computer
  Graphics} \bibinfo{volume}{24} (\bibinfo{year}{2018})
  \bibinfo{pages}{298--308}. \DOIprefix\doi{10.1109/TVCG.2017.2744818}.
\bibitem[{Roesch(1999)}]{Roesch1999}
\bibinfo{author}{M.~Roesch},
\newblock \bibinfo{title}{{SNORT} - lightweight intrusion detection for
  networks},
\newblock in: \bibinfo{booktitle}{Proceedings of the 13th USENIX conference on
  System administration}, LISA '99, \bibinfo{publisher}{USENIX Association},
  \bibinfo{address}{Berkeley, CA, USA}, \bibinfo{year}{1999}, pp.
  \bibinfo{pages}{229--238}. \bibinfo{note}{ISBN 978-1-931971-59-1}.
\bibitem[{Paxson(1999)}]{Paxson1999}
\bibinfo{author}{V.~Paxson},
\newblock \bibinfo{title}{{BRO}: a system for detecting network intruders in
  real-time},
\newblock \bibinfo{journal}{Computer Networks} \bibinfo{volume}{31}
  (\bibinfo{year}{1999}) \bibinfo{pages}{2435--2463}.
\bibitem[{Lee and Stolfo(1998)}]{Lee1998}
\bibinfo{author}{W.~Lee}, \bibinfo{author}{S.~J. Stolfo},
\newblock \bibinfo{title}{Data mining approaches for intrusion detection},
\newblock in: \bibinfo{booktitle}{Proceedings of the 7th conference on USENIX
  Security Symposium - Volume 7}, \bibinfo{publisher}{USENIX Association},
  \bibinfo{address}{Berkeley, CA, USA}, \bibinfo{year}{1998}, pp.
  \bibinfo{pages}{6--6}. \bibinfo{note}{ISBN 978-1-931971-59-1}.
\bibitem[{Lippmann et~al.(2000)Lippmann, Fried, Graf, Haines, Kendall, McClung,
  Weber, Webster, Wyschogrod, Cunningham, and Zissman}]{Lippmann2000}
\bibinfo{author}{R.~P. Lippmann}, \bibinfo{author}{D.~J. Fried},
  \bibinfo{author}{I.~Graf}, \bibinfo{author}{J.~W. Haines},
  \bibinfo{author}{K.~R. Kendall}, \bibinfo{author}{D.~McClung},
  \bibinfo{author}{D.~Weber}, \bibinfo{author}{S.~E. Webster},
  \bibinfo{author}{D.~Wyschogrod}, \bibinfo{author}{R.~K. Cunningham},
  \bibinfo{author}{M.~A. Zissman},
\newblock \bibinfo{title}{{Evaluating intrusion detection systems: The 1998
  DARPA off-line intrusion detection evaluation}},
\newblock \bibinfo{journal}{Proceedings - DARPA Information Survivability
  Conference and Exposition, DISCEX 2000} \bibinfo{volume}{2}
  (\bibinfo{year}{2000}) \bibinfo{pages}{12--26}.
  \DOIprefix\doi{10.1109/DISCEX.2000.821506}.
\bibitem[{Kenyon et~al.(2020)Kenyon, Deka, and Elizondo}]{Kenyon2020}
\bibinfo{author}{A.~Kenyon}, \bibinfo{author}{L.~Deka},
  \bibinfo{author}{D.~Elizondo},
\newblock \bibinfo{title}{Are public intrusion datasets fit for purpose
  characterising the state of the art in intrusion event datasets},
\newblock \bibinfo{journal}{Computers \& Security} \bibinfo{volume}{99}
  (\bibinfo{year}{2020}) \bibinfo{pages}{102022}.
  \DOIprefix\doi{https://doi.org/10.1016/j.cose.2020.102022}.
\bibitem[{Ugarte-Pedrero et~al.(2019)Ugarte-Pedrero, Graziano, and
  Balzarotti}]{ugarte2019}
\bibinfo{author}{X.~Ugarte-Pedrero}, \bibinfo{author}{M.~Graziano},
  \bibinfo{author}{D.~Balzarotti},
\newblock \bibinfo{title}{A close look at a daily dataset of malware samples},
\newblock \bibinfo{journal}{ACM Trans. Priv. Secur.} \bibinfo{volume}{22}
  (\bibinfo{year}{2019}). \URLprefix \url{https://doi.org/10.1145/3291061}.
  \DOIprefix\doi{10.1145/3291061}.
\bibitem[{Garc{\'{i}}a et~al.(2014)Garc{\'{i}}a, Grill, Stiborek, and
  Zunino}]{Garcia2014}
\bibinfo{author}{S.~Garc{\'{i}}a}, \bibinfo{author}{M.~Grill},
  \bibinfo{author}{J.~Stiborek}, \bibinfo{author}{A.~Zunino},
\newblock \bibinfo{title}{{An empirical comparison of botnet detection
  methods}},
\newblock \bibinfo{journal}{Computers and Security} \bibinfo{volume}{45}
  (\bibinfo{year}{2014}) \bibinfo{pages}{100--123}.
  \DOIprefix\doi{10.1016/j.cose.2014.05.011}.
\bibitem[{Papadogiannaki and Ioannidis(2021)}]{Papadogiannaki2021}
\bibinfo{author}{E.~Papadogiannaki}, \bibinfo{author}{S.~Ioannidis},
\newblock \bibinfo{title}{A survey on encrypted network traffic analysis
  applications, techniques, and countermeasures},
\newblock \bibinfo{journal}{ACM Comput. Surv.} \bibinfo{volume}{54}
  (\bibinfo{year}{2021}). \URLprefix \url{https://doi.org/10.1145/3457904}.
  \DOIprefix\doi{10.1145/3457904}.
\bibitem[{Bhuyan et~al.(2015)Bhuyan, Bhattacharyya, and Kalita}]{Bhuyan2015}
\bibinfo{author}{M.~H. Bhuyan}, \bibinfo{author}{D.~K. Bhattacharyya},
  \bibinfo{author}{J.~K. Kalita},
\newblock \bibinfo{title}{{Towards generating real-life datasets for network
  intrusion detection}},
\newblock \bibinfo{journal}{International Journal of Network Security}
  \bibinfo{volume}{17} (\bibinfo{year}{2015}) \bibinfo{pages}{683--701}.
\bibitem[{Moustafa and Slay(2015)}]{Moustafa2015}
\bibinfo{author}{N.~Moustafa}, \bibinfo{author}{J.~Slay},
\newblock \bibinfo{title}{{UNSW-NB15: A comprehensive data set for network
  intrusion detection systems (UNSW-NB15 network data set)}},
\newblock \bibinfo{journal}{2015 Military Communications and Information
  Systems Conference, MilCIS 2015 - Proceedings}  (\bibinfo{year}{2015}).
  \DOIprefix\doi{10.1109/MilCIS.2015.7348942}.
\bibitem[{Haider et~al.(2017)Haider, Hu, Slay, Turnbull, and Xie}]{Haider2017}
\bibinfo{author}{W.~Haider}, \bibinfo{author}{J.~Hu},
  \bibinfo{author}{J.~Slay}, \bibinfo{author}{B.~P. Turnbull},
  \bibinfo{author}{Y.~Xie},
\newblock \bibinfo{title}{{Generating realistic intrusion detection system
  dataset based on fuzzy qualitative modeling}},
\newblock \bibinfo{journal}{Journal of Network and Computer Applications}
  \bibinfo{volume}{87} (\bibinfo{year}{2017}) \bibinfo{pages}{185--192}.
  \DOIprefix\doi{10.1016/j.jnca.2017.03.018}.
\bibitem[{Mukkavilli et~al.(2016)Mukkavilli, Shetty, and Hong}]{Mukkavilli2016}
\bibinfo{author}{S.~K. Mukkavilli}, \bibinfo{author}{S.~Shetty},
  \bibinfo{author}{L.~Hong},
\newblock \bibinfo{title}{{Generation of Labelled Datasets to Quantify the
  Impact of Security Threats to Cloud Data Centers}},
\newblock \bibinfo{journal}{Journal of Information Security}
  (\bibinfo{year}{2016}) \bibinfo{pages}{172--184}.
\bibitem[{Lemay and Fernandez(2016)}]{Lemay2016}
\bibinfo{author}{A.~Lemay}, \bibinfo{author}{J.~M. Fernandez},
\newblock \bibinfo{title}{{Providing SCADA network data sets for intrusion
  detection research}},
\newblock \bibinfo{journal}{Usenix Cset}  (\bibinfo{year}{2016}).
\bibitem[{Shi(????)}]{Shiravi2012}
  (????).
\bibitem[{Stolfo et~al.(2000)Stolfo, Fan, Lee, Prodromidis, and
  Chan}]{Stolfo2000}
\bibinfo{author}{S.~J. Stolfo}, \bibinfo{author}{W.~Fan},
  \bibinfo{author}{W.~Lee}, \bibinfo{author}{A.~Prodromidis},
  \bibinfo{author}{P.~K. Chan},
\newblock \bibinfo{title}{{Cost-based modeling for fraud and intrusion
  detection: Results from the JAM project}},
\newblock \bibinfo{journal}{Proceedings - DARPA Information Survivability
  Conference and Exposition, DISCEX 2000} \bibinfo{volume}{2}
  (\bibinfo{year}{2000}) \bibinfo{pages}{130--144}.
  \DOIprefix\doi{10.1109/DISCEX.2000.821515}.
\bibitem[{Catania et~al.(2012)Catania, Bromberg, and
  GarciaGarino}]{Catania2012b}
\bibinfo{author}{C.~A. Catania}, \bibinfo{author}{F.~Bromberg},
  \bibinfo{author}{C.~GarciaGarino},
\newblock \bibinfo{title}{An autonomous labeling approach to support vector
  machines algorithms for network traffic anomaly detection},
\newblock \bibinfo{journal}{Expert Syst. Appl.} \bibinfo{volume}{39}
  (\bibinfo{year}{2012}) \bibinfo{pages}{1822–1829}. \URLprefix
  \url{https://doi.org/10.1016/j.eswa.2011.08.068}.
  \DOIprefix\doi{10.1016/j.eswa.2011.08.068}.
\bibitem[{Gargiulo et~al.(2012)Gargiulo, Mazzariello, and
  Sansone}]{Gargiulo2012}
\bibinfo{author}{F.~Gargiulo}, \bibinfo{author}{C.~Mazzariello},
  \bibinfo{author}{C.~Sansone},
\newblock \bibinfo{title}{{Automatically building datasets of labeled IP
  traffic traces: A self-training approach}},
\newblock \bibinfo{journal}{Applied Soft Computing Journal}
  \bibinfo{volume}{12} (\bibinfo{year}{2012}) \bibinfo{pages}{1640--1649}.
  \DOIprefix\doi{10.1016/j.asoc.2012.02.012}.
\bibitem[{Song et~al.(2011)Song, Takakura, Okabe, Eto, Inoue, and
  Nakao}]{Song2011}
\bibinfo{author}{J.~Song}, \bibinfo{author}{H.~Takakura},
  \bibinfo{author}{Y.~Okabe}, \bibinfo{author}{M.~Eto},
  \bibinfo{author}{D.~Inoue}, \bibinfo{author}{K.~Nakao},
\newblock \bibinfo{title}{{Statistical analysis of honeypot data and building
  of Kyoto 2006+ dataset for NIDS evaluation}},
\newblock \bibinfo{journal}{Proceedings of the 1st Workshop on Building
  Analysis Datasets and Gathering Experience Returns for Security, BADGERS
  2011}  (\bibinfo{year}{2011}) \bibinfo{pages}{29--36}.
  \DOIprefix\doi{10.1145/1978672.1978676}.
\bibitem[{Sperotto et~al.(2009)Sperotto, Sadre, van Vliet, and
  Pras}]{Sperotto2009}
\bibinfo{author}{A.~Sperotto}, \bibinfo{author}{R.~Sadre},
  \bibinfo{author}{F.~van Vliet}, \bibinfo{author}{A.~Pras},
\newblock \bibinfo{title}{A labeled data set for flow-based intrusion
  detection},
\newblock in: \bibinfo{editor}{G.~Nunzi}, \bibinfo{editor}{C.~Scoglio},
  \bibinfo{editor}{X.~Li} (Eds.), \bibinfo{booktitle}{IP Operations and
  Management}, \bibinfo{publisher}{Springer Berlin Heidelberg},
  \bibinfo{address}{Berlin, Heidelberg}, \bibinfo{year}{2009}, pp.
  \bibinfo{pages}{39--50}.
\bibitem[{Aparicio-Navarro et~al.(2014)Aparicio-Navarro, Kyriakopoulos, and
  Parish}]{Aparicio-Navarro2014}
\bibinfo{author}{F.~J. Aparicio-Navarro}, \bibinfo{author}{K.~G.
  Kyriakopoulos}, \bibinfo{author}{D.~J. Parish},
\newblock \bibinfo{title}{{Automatic dataset labelling and feature selection
  for intrusion detection systems}},
\newblock \bibinfo{journal}{Proceedings - IEEE Military Communications
  Conference MILCOM}  (\bibinfo{year}{2014}) \bibinfo{pages}{46--51}.
  \DOIprefix\doi{10.1109/MILCOM.2014.17}.
\bibitem[{Ring et~al.(2017)Ring, Wunderlich, Grüdl, Landes, and
  Hotho}]{Ring2017361}
\bibinfo{author}{M.~Ring}, \bibinfo{author}{S.~Wunderlich},
  \bibinfo{author}{D.~Grüdl}, \bibinfo{author}{D.~Landes},
  \bibinfo{author}{A.~Hotho},
\newblock \bibinfo{title}{Flow-based benchmark data sets for intrusion
  detection},
\newblock \bibinfo{year}{2017}, pp. \bibinfo{pages}{361--369}.
  \bibinfo{note}{Cited By 54}.
\bibitem[{Sangster et~al.(2009)Sangster, Cook, Fanelli, Dean, Adams, Morrell,
  and Conti}]{Sangster2009}
\bibinfo{author}{B.~Sangster}, \bibinfo{author}{T.~Cook},
  \bibinfo{author}{R.~Fanelli}, \bibinfo{author}{E.~Dean},
  \bibinfo{author}{W.~J. Adams}, \bibinfo{author}{C.~Morrell},
  \bibinfo{author}{G.~Conti},
\newblock \bibinfo{title}{{Toward Instrumenting Network Warfare Competitions to
  Generate Labeled Datasets}},
\newblock \bibinfo{journal}{USENIX Security's Workshop on Cyber Security
  Experimentation and Test (CSET)}  (\bibinfo{year}{2009}).
\bibitem[{University(2007)}]{PlanetLab}
\bibinfo{author}{P.~University}, \bibinfo{title}{Planet lab},
  \bibinfo{howpublished}{http://www.https://www.planet-lab.org/status},
  \bibinfo{year}{2007}. \bibinfo{note}{[Online; accessed January-2020]}.
\bibitem[{Peterson et~al.(2006)Peterson, Bavier, Fiuczynski, and
  Muir}]{planetlab2006}
\bibinfo{author}{L.~Peterson}, \bibinfo{author}{A.~Bavier},
  \bibinfo{author}{M.~E. Fiuczynski}, \bibinfo{author}{S.~Muir},
\newblock \bibinfo{title}{Experiences building planetlab},
\newblock in: \bibinfo{booktitle}{Proceedings of the 7th Symposium on Operating
  Systems Design and Implementation}, OSDI '06, \bibinfo{publisher}{USENIX
  Association}, \bibinfo{address}{USA}, \bibinfo{year}{2006}, p.
  \bibinfo{pages}{351–366}.
\bibitem[{Shiravi et~al.(2012)Shiravi, Shiravi, and
  Ghorbani}]{Shiravi-Survey2012}
\bibinfo{author}{H.~Shiravi}, \bibinfo{author}{A.~Shiravi},
  \bibinfo{author}{A.~A. Ghorbani},
\newblock \bibinfo{title}{{A survey of visualization systems for network
  security}},
\newblock \bibinfo{journal}{IEEE Transactions on Visualization and Computer
  Graphics} \bibinfo{volume}{18} (\bibinfo{year}{2012})
  \bibinfo{pages}{1313--1329}. \DOIprefix\doi{10.1109/TVCG.2011.144}.
\bibitem[{Huitsing et~al.(2008)Huitsing, Chandia, Papa, and
  Shenoi}]{modbus2008}
\bibinfo{author}{P.~Huitsing}, \bibinfo{author}{R.~Chandia},
  \bibinfo{author}{M.~Papa}, \bibinfo{author}{S.~Shenoi},
\newblock \bibinfo{title}{Attack taxonomies for the modbus protocols},
\newblock \bibinfo{journal}{International Journal of Critical Infrastructure
  Protection} \bibinfo{volume}{1} (\bibinfo{year}{2008})
  \bibinfo{pages}{37--44}.
\bibitem[{Lippmann et~al.(2000)Lippmann, Haines, Fried, Korba, and
  Das}]{Lippmann2000b}
\bibinfo{author}{R.~Lippmann}, \bibinfo{author}{J.~W. Haines},
  \bibinfo{author}{D.~J. Fried}, \bibinfo{author}{J.~Korba},
  \bibinfo{author}{K.~Das},
\newblock \bibinfo{title}{The 1999 darpa off-line intrusion detection
  evaluation},
\newblock \bibinfo{journal}{Computer Networks} \bibinfo{volume}{34}
  (\bibinfo{year}{2000}) \bibinfo{pages}{579--595}.
  \DOIprefix\doi{https://doi.org/10.1016/S1389-1286(00)00139-0},
  \bibinfo{note}{recent Advances in Intrusion Detection Systems}.
\bibitem[{Stokes et~al.(2008)Stokes, Platt, Kravis, and Shilman}]{aladin}
\bibinfo{author}{J.~Stokes}, \bibinfo{author}{J.~Platt},
  \bibinfo{author}{J.~Kravis}, \bibinfo{author}{M.~Shilman},
  \bibinfo{title}{ALADIN: Active Learning of Anomalies to Detect Intrusions},
  \bibinfo{type}{Technical Report} \bibinfo{number}{MSR-TR-2008-24},
  \bibinfo{year}{2008}.
\bibitem[{Beaugnon et~al.(2012)Beaugnon, Chifflier, and Bach}]{Chifflier2012}
\bibinfo{author}{A.~Beaugnon}, \bibinfo{author}{P.~Chifflier},
  \bibinfo{author}{F.~Bach},
\newblock \bibinfo{title}{{ILAB: An Interactive Labelling Strategy for
  Intrusion Detection}},
\newblock \bibinfo{journal}{International Symposium on Research in Attacks,
  Intrusions, and Defenses} \bibinfo{volume}{7462} (\bibinfo{year}{2012})
  \bibinfo{pages}{120--140}. \DOIprefix\doi{10.1007/978-3-642-33338-5}.
  \href{http://arxiv.org/abs/9780201398298}{{\tt arXiv:9780201398298}}.
\bibitem[{Fan et~al.(2019)Fan, Li, Yuan, Dong, and Liang}]{Fan2019}
\bibinfo{author}{X.~Fan}, \bibinfo{author}{C.~Li}, \bibinfo{author}{X.~Yuan},
  \bibinfo{author}{X.~Dong}, \bibinfo{author}{J.~Liang},
\newblock \bibinfo{title}{{An interactive visual analytics approach for network
  anomaly detection through smart labeling}},
\newblock \bibinfo{journal}{Journal of Visualization} \bibinfo{volume}{22}
  (\bibinfo{year}{2019}) \bibinfo{pages}{955--971}.
  \DOIprefix\doi{10.1007/s12650-019-00580-7}.
\bibitem[{G{\"{o}}rnitz et~al.(2013)G{\"{o}}rnitz, Kloft, Rieck, and
  Brefeld}]{Gornitz2013}
\bibinfo{author}{N.~G{\"{o}}rnitz}, \bibinfo{author}{M.~Kloft},
  \bibinfo{author}{K.~Rieck}, \bibinfo{author}{U.~Brefeld},
\newblock \bibinfo{title}{{Toward Supervised Anomaly Detection}},
\newblock \bibinfo{journal}{Journal of Artificial Intelligence Research}
  \bibinfo{volume}{46} (\bibinfo{year}{2013}) \bibinfo{pages}{235--262}.
  \DOIprefix\doi{10.1613/jair.3623}.
\bibitem[{Koike et~al.(2006)Koike, Ohno, and Koizumi}]{Koike2006}
\bibinfo{author}{H.~Koike}, \bibinfo{author}{K.~Ohno},
  \bibinfo{author}{K.~Koizumi},
\newblock \bibinfo{title}{{Visualizing Cyber Attacks using IP Matrix}},
\newblock \bibinfo{journal}{IEEE Workshop on Visualization for Computer
  Security}  (\bibinfo{year}{2006}) \bibinfo{pages}{11--11}.
  \DOIprefix\doi{10.1109/vizsec.2005.22}.
\bibitem[{Livnat et~al.(2005)Livnat, Agutter, Moon, Erbacher, and
  Foresti}]{Livnat2005}
\bibinfo{author}{Y.~Livnat}, \bibinfo{author}{J.~Agutter},
  \bibinfo{author}{S.~Moon}, \bibinfo{author}{R.~F. Erbacher},
  \bibinfo{author}{S.~Foresti},
\newblock \bibinfo{title}{{A visualization paradigm for network intrusion
  detection}},
\newblock \bibinfo{journal}{Proceedings from the 6th Annual IEEE System, Man
  and Cybernetics Information Assurance Workshop, SMC 2005}
  \bibinfo{volume}{2005} (\bibinfo{year}{2005}) \bibinfo{pages}{92--99}.
  \DOIprefix\doi{10.1109/IAW.2005.1495939}.
\bibitem[{Ren et~al.(2005)Ren, Gao, Li, Chen, and Watson}]{Ren2005}
\bibinfo{author}{P.~Ren}, \bibinfo{author}{Y.~Gao}, \bibinfo{author}{Z.~Li},
  \bibinfo{author}{Y.~Chen}, \bibinfo{author}{B.~Watson},
\newblock \bibinfo{title}{{IDGraphs: Intrusion detection and analysis using
  histographs}},
\newblock \bibinfo{journal}{IEEE Workshop on Visualization for Computer
  Security 2005, VizSEC 05, Proceedings}  (\bibinfo{year}{2005})
  \bibinfo{pages}{39--46}. \DOIprefix\doi{10.1109/VIZSEC.2005.1532064}.
\bibitem[{Scott et~al.(2003)Scott, Nyarko, Capers, and
  Ladeji-Osias}]{Scott2003}
\bibinfo{author}{C.~Scott}, \bibinfo{author}{K.~Nyarko},
  \bibinfo{author}{T.~Capers}, \bibinfo{author}{J.~Ladeji-Osias},
\newblock \bibinfo{title}{{Network intrusion visualization with niva, an
  intrusion detection visual and haptic analyzer}},
\newblock \bibinfo{journal}{Information Visualization} \bibinfo{volume}{2}
  (\bibinfo{year}{2003}) \bibinfo{pages}{82--94}.
  \DOIprefix\doi{10.1057/palgrave.ivs.9500044}.
\bibitem[{Chen et~al.(2014)Chen, Guo, Yuan, Merkle, Schaefer, and
  Ertl}]{Chen2014}
\bibinfo{author}{S.~Chen}, \bibinfo{author}{C.~Guo}, \bibinfo{author}{X.~Yuan},
  \bibinfo{author}{F.~Merkle}, \bibinfo{author}{H.~Schaefer},
  \bibinfo{author}{T.~Ertl},
\newblock \bibinfo{title}{Oceans - online collaborative explorative analysis on
  network security},
\newblock volume \bibinfo{volume}{10-November-2014},
  \bibinfo{publisher}{Association for Computing Machinery},
  \bibinfo{year}{2014}, pp. \bibinfo{pages}{1--8}.
  \DOIprefix\doi{10.1145/2671491.2671493}.
\bibitem[{{Cappers} et~al.(2018){Cappers}, {Meessen}, {Etalle}, and {van
  Wijk}}]{Cappers2018}
\bibinfo{author}{B.~C.~M. {Cappers}}, \bibinfo{author}{P.~N. {Meessen}},
  \bibinfo{author}{S.~{Etalle}}, \bibinfo{author}{J.~J. {van Wijk}},
\newblock \bibinfo{title}{Eventpad: Rapid malware analysis and reverse
  engineering using visual analytics},
\newblock in: \bibinfo{booktitle}{2018 IEEE Symposium on Visualization for
  Cyber Security (VizSec)}, \bibinfo{year}{2018}, pp. \bibinfo{pages}{1--8}.
\bibitem[{Zhang et~al.(2019)Zhang, Guo, Li, and Angin}]{Zhang2019}
\bibinfo{author}{H.~Zhang}, \bibinfo{author}{Y.~Guo}, \bibinfo{author}{T.~Li},
  \bibinfo{author}{P.~Angin},
\newblock \bibinfo{title}{Multifeature named entity recognition in information
  security based on adversarial learning},
\newblock \bibinfo{journal}{Security and Communication Networks}
  \bibinfo{volume}{2019} (\bibinfo{year}{2019}).
  \DOIprefix\doi{10.1155/2019/6417407}.
\bibitem[{Shahid et~al.(2018)Shahid, Blanc, Zhang, and Debar}]{Shahid2018}
\bibinfo{author}{M.~R. Shahid}, \bibinfo{author}{G.~Blanc},
  \bibinfo{author}{Z.~Zhang}, \bibinfo{author}{H.~Debar},
\newblock \bibinfo{title}{Iot devices recognition through network traffic
  analysis},
\newblock in: \bibinfo{booktitle}{2018 IEEE International Conference on Big
  Data (Big Data)}, \bibinfo{year}{2018}, pp. \bibinfo{pages}{5187--5192}.
  \DOIprefix\doi{10.1109/BigData.2018.8622243}.
\bibitem[{Yang et~al.(2015)Yang, Ma, Nie, Chang, and Hauptmann}]{Yang2015}
\bibinfo{author}{Y.~Yang}, \bibinfo{author}{Z.~Ma}, \bibinfo{author}{F.~Nie},
  \bibinfo{author}{X.~Chang}, \bibinfo{author}{A.~G. Hauptmann},
\newblock \bibinfo{title}{{Multi-Class Active Learning by Uncertainty Sampling
  with Diversity Maximization}},
\newblock \bibinfo{journal}{International Journal of Computer Vision}
  \bibinfo{volume}{113} (\bibinfo{year}{2015}) \bibinfo{pages}{113--127}.
  \DOIprefix\doi{10.1007/s11263-014-0781-x}.
\bibitem[{McElwee(2017)}]{McElwee2017}
\bibinfo{author}{S.~McElwee},
\newblock \bibinfo{title}{Active learning intrusion detection using k-means
  clustering selection},
\newblock in: \bibinfo{booktitle}{SoutheastCon 2017}, \bibinfo{year}{2017}, pp.
  \bibinfo{pages}{1--7}. \DOIprefix\doi{10.1109/SECON.2017.7925383}.
\bibitem[{Lewis and Catlett(1994)}]{LEWIS1994148}
\bibinfo{author}{D.~D. Lewis}, \bibinfo{author}{J.~Catlett},
\newblock \bibinfo{title}{Heterogeneous uncertainty sampling for supervised
  learning},
\newblock in: \bibinfo{editor}{W.~W. Cohen}, \bibinfo{editor}{H.~Hirsh} (Eds.),
  \bibinfo{booktitle}{Machine Learning Proceedings 1994},
  \bibinfo{publisher}{Morgan Kaufmann}, \bibinfo{address}{San Francisco (CA)},
  \bibinfo{year}{1994}, pp. \bibinfo{pages}{148--156}.
  \DOIprefix\doi{https://doi.org/10.1016/B978-1-55860-335-6.50026-X}.
\bibitem[{Almgren and Jonsson(2004)}]{Almgren2004}
\bibinfo{author}{M.~Almgren}, \bibinfo{author}{E.~Jonsson},
\newblock \bibinfo{title}{{Using active learning in intrusion detection}},
\newblock in: \bibinfo{booktitle}{Proceedings of the Computer Security
  Foundations Workshop}, volume~\bibinfo{volume}{17}, \bibinfo{year}{2004}, pp.
  \bibinfo{pages}{88--98}. \DOIprefix\doi{10.1109/csfw.2004.1310734}.
\bibitem[{Beaugnon et~al.(2017)Beaugnon, Chifflier, and Bach}]{Beaugnon2017}
\bibinfo{author}{A.~Beaugnon}, \bibinfo{author}{P.~Chifflier},
  \bibinfo{author}{F.~Bach},
\newblock \bibinfo{title}{Ilab: An interactive labelling strategy for intrusion
  detection},
\newblock in: \bibinfo{editor}{M.~Dacier}, \bibinfo{editor}{M.~Bailey},
  \bibinfo{editor}{M.~Polychronakis}, \bibinfo{editor}{M.~Antonakakis} (Eds.),
  \bibinfo{booktitle}{Research in Attacks, Intrusions, and Defenses},
  \bibinfo{publisher}{Springer International Publishing},
  \bibinfo{address}{Cham}, \bibinfo{year}{2017}, pp. \bibinfo{pages}{120--140}.
\bibitem[{{Guerra} et~al.(2019){Guerra}, {Veas}, and
  {Catania}}]{guerra2019vizsec}
\bibinfo{author}{J.~L. {Guerra}}, \bibinfo{author}{E.~{Veas}},
  \bibinfo{author}{C.~A. {Catania}},
\newblock \bibinfo{title}{A study on labeling network hostile behavior with
  intelligent interactive tools},
\newblock in: \bibinfo{booktitle}{2019 IEEE Symposium on Visualization for
  Cyber Security (VizSec)}, \bibinfo{year}{2019}, pp. \bibinfo{pages}{1--10}.
  \DOIprefix\doi{10.1109/VizSec48167.2019.9161489}.
\bibitem[{Pelleg and Moore(2004)}]{Pelleg2004}
\bibinfo{author}{D.~Pelleg}, \bibinfo{author}{A.~Moore},
\newblock \bibinfo{title}{{Active learning for anomaly and rare-category
  detection}},
\newblock \bibinfo{journal}{Advances in Neural Information Processing Systems}
  \bibinfo{volume}{18} (\bibinfo{year}{2004}) \bibinfo{pages}{1073--1080}.
\bibitem[{Torres et~al.(2019)Torres, Catania, and Veas}]{Guerra2019}
\bibinfo{author}{J.~L. Torres}, \bibinfo{author}{C.~A. Catania},
  \bibinfo{author}{E.~Veas},
\newblock \bibinfo{title}{{Active learning approach to label network traffic
  datasets}},
\newblock \bibinfo{journal}{Journal of Information Security and Applications}
  \bibinfo{volume}{49} (\bibinfo{year}{2019}) \bibinfo{pages}{102388}.
  \DOIprefix\doi{10.1016/j.jisa.2019.102388}.
\bibitem[{Hofstede et~al.(2018)Hofstede, Pras, Sperotto, and
  Rodosek}]{Hofstede2018}
\bibinfo{author}{R.~Hofstede}, \bibinfo{author}{A.~Pras},
  \bibinfo{author}{A.~Sperotto}, \bibinfo{author}{G.~D. Rodosek},
\newblock \bibinfo{title}{Flow-based compromise detection: Lessons learned},
\newblock \bibinfo{journal}{IEEE Security Privacy} \bibinfo{volume}{16}
  (\bibinfo{year}{2018}) \bibinfo{pages}{82--89}.
  \DOIprefix\doi{10.1109/MSP.2018.1331021}.
\bibitem[{Cermak et~al.(2018)Cermak, Jirsik, Velan, Komarkova, Spacek, Drasar,
  and Plesnik}]{Cermak2018}
\bibinfo{author}{M.~Cermak}, \bibinfo{author}{T.~Jirsik},
  \bibinfo{author}{P.~Velan}, \bibinfo{author}{J.~Komarkova},
  \bibinfo{author}{S.~Spacek}, \bibinfo{author}{M.~Drasar},
  \bibinfo{author}{T.~Plesnik},
\newblock \bibinfo{title}{Towards provable network traffic measurement and
  analysis via semi-labeled trace datasets},
\newblock in: \bibinfo{booktitle}{2018 Network Traffic Measurement and Analysis
  Conference (TMA)}, \bibinfo{year}{2018}, pp. \bibinfo{pages}{1--8}.
  \DOIprefix\doi{10.23919/TMA.2018.8506498}.
\bibitem[{Guerra et~al.(2017)Guerra, Catania, and Veas}]{Guerra2017}
\bibinfo{author}{J.~Guerra}, \bibinfo{author}{C.~A. Catania},
  \bibinfo{author}{E.~Veas},
\newblock \bibinfo{title}{{Visual Exploration of Network Hostile Behavior}},
\newblock \bibinfo{journal}{Proceedings of the 2017 ACM Workshop on Exploratory
  Search and Interactive Data Analytics - ESIDA '17}  (\bibinfo{year}{2017})
  \bibinfo{pages}{51--54}. \DOIprefix\doi{10.1145/3038462.3038466}.
\bibitem[{Nehinbe(2011)}]{Nehinbe2011}
\bibinfo{author}{J.~O. Nehinbe},
\newblock \bibinfo{title}{{A critical evaluation of datasets for investigating
  IDSs and IPSs researches}},
\newblock \bibinfo{journal}{Proceedings of 2011, 10th IEEE International
  Conference on Cybernetic Intelligent Systems, CIS 2011}
  (\bibinfo{year}{2011}) \bibinfo{pages}{92--97}.
  \DOIprefix\doi{10.1109/CIS.2011.6169141}.
\bibitem[{Wang and Jones(2021)}]{Wang2021}
\bibinfo{author}{L.~Wang}, \bibinfo{author}{R.~Jones},
\newblock \bibinfo{title}{Big data analytics in cyber security: Network traffic
  and attacks},
\newblock \bibinfo{journal}{Journal of Computer Information Systems}
  \bibinfo{volume}{61} (\bibinfo{year}{2021}) \bibinfo{pages}{410--417}.
  \URLprefix \url{https://doi.org/10.1080/08874417.2019.1688731}.
  \DOIprefix\doi{10.1080/08874417.2019.1688731}.
  \href{http://arxiv.org/abs/https://doi.org/10.1080/08874417.2019.1688731}{{\tt
  arXiv:https://doi.org/10.1080/08874417.2019.1688731}}.
\bibitem[{Banerjee et~al.(2020)Banerjee, Chen, Hung, Huang, Wang, and
  Chevesaran}]{banerjee2020challenges}
\bibinfo{author}{A.~Banerjee}, \bibinfo{author}{C.-C. Chen},
  \bibinfo{author}{C.-C. Hung}, \bibinfo{author}{X.~Huang},
  \bibinfo{author}{Y.~Wang}, \bibinfo{author}{R.~Chevesaran},
\newblock \bibinfo{title}{Challenges and experiences with mlops for performance
  diagnostics in hybrid-cloud enterprise software deployments},
\newblock in: \bibinfo{booktitle}{2020 $\{$USENIX$\}$ Conference on Operational
  Machine Learning (OpML 20)}, \bibinfo{year}{2020}.
\bibitem[{Gharib et~al.(2016)Gharib, Sharafaldin†, Lashkari, and
  Ghorbani}]{Gharib2016}
\bibinfo{author}{A.~Gharib}, \bibinfo{author}{I.~Sharafaldin†},
  \bibinfo{author}{A.~H. Lashkari}, \bibinfo{author}{A.~A. Ghorbani},
\newblock \bibinfo{title}{{An Evaluation Framework for Intrusion Detection
  Dataset}},
\newblock \bibinfo{journal}{International Conference on Information Science and
  Security (ICISS)} \bibinfo{volume}{22} (\bibinfo{year}{2016})
  \bibinfo{pages}{1--6}. \DOIprefix\doi{10.1016/0371-1951(66)80211-4}.

\end{thebibliography}







\end{document}